\documentclass[a4paper,11pt]{article}
\pdfoutput=1 

\usepackage{jheppub} 

\usepackage[T1]{fontenc} 

\def\hbbj{$H\rightarrow b\overline{b}j \;$}
\def\hbb{$H\rightarrow b\overline{b}\;$}

\def\spa#1.#2{\langle #1 #2\rangle}
\def\spb#1.#2{[ #1 #2]}
\def\spab#1.#2.#3{\langle #1 |#2| #3] }
\def\nnnlo{N$^{3}$LO\;}
\def\nnlo{NNLO\;}
\def\nlo{NLO\;}

\title{\boldmath \nnnlo  predictions for the decay of the Higgs boson to bottom quarks}

\author{Roberto Mondini,}         
\author{Matthew Schiavi,}       
\author{and Ciaran Williams}    

\affiliation{Department of Physics,\\University at Buffalo, The State University of New York, Buffalo
14260, USA}

\emailAdd{rmondini@buffalo.edu}
\emailAdd{mmschiav@buffalo.edu}
\emailAdd{ciaranwi@buffalo.edu}

\begin{abstract}{We present a fully-differential calculation of the $H\rightarrow b\overline{b}$ decay at next-to-next-to-next-to-leading order (N$^{3}$LO) accuracy. Our calculation considers diagrams in which the Higgs boson couples directly to the bottom quarks, i.e.~the perturbative order we consider is $\mathcal{O}(\alpha_s^3 y_b^2)$. In order to regulate the infrared divergences present at this order we use  
the Projection-to-Born technique coupled with $N$-jettiness slicing.  After validating our methodology at next-to-next-to-leading order (NNLO) we present exclusive jet rates and differential distributions for jet observables at \nnnlo accuracy using the Durham jet algorithm in the Higgs rest frame. }
\end{abstract}

\begin{document} 
\maketitle
\flushbottom

\section{Introduction}

The discovery of a Higgs boson~\cite{Aad:2012tfa,Chatrchyan:2012xdj} at CERN's Large Hadron Collider (LHC) represents the most significant result in high energy physics in recent history. 
Over the next couple of decades continued measurements of the properties of the Higgs will result in increasingly-stringent tests of the predictions from the Standard Model (SM). 
These studies will continue to take place at the LHC (including the future high-luminosity upgrade) and putative future colliders, which are currently in the early design phases \cite{Gomez-Ceballos:2013zzn,Baer:2013cma,Benedikt:2018qee}. From a Higgs precision viewpoint, one strongly-motivated future accelerator is a lepton collider, capable of producing a large data set with small experimental uncertainties and thus allowing precision studies 
of the Higgs boson akin to what was successfully performed at LEP for the $Z$ boson. In order to achieve these goals, it is vital for the theoretical community to provide precise predictions for Higgs-related observables with accuracies at the few-percent to per-mille level. 

For the 125-GeV Higgs boson the predominant decay mode is to a pair of bottom quarks ($b\overline{b}$), whose partial width accounts for around 60\% of the total. An accurate measurement of \hbb is therefore crucial, since the Higgs-bottom Yukawa coupling ($y_b$) enters every LHC Higgs measurement through the total width. In a hadronic environment the measurement of \hbb is particularly challenging due to the presence of large QCD backgrounds. In order to overcome these obstacles, experimental analyses typically focus on associated ($VH$) production modes, which have more manageable backgrounds \cite{Aaboud:2018zhk,Sirunyan:2018kst}. However, using jet-substructure techniques it is also possible to access \hbb through the gluon-fusion production mode (at high transverse momenta)~\cite{Sirunyan:2017dgcx}.

Given its importance to Higgs physics, the \hbb decay has been studied in the literature for many years~\cite{Braaten:1980yq,Gorishnii:1990zu,Kataev:1993be,Surguladze:1994gc,Larin:1995sq,Chetyrkin:1995pd,Chetyrkin:1996sr}. Currently, higher-order corrections from QCD are known up to N$^4$LO (i.e.~up to order $\mathcal{O}(\alpha_s^4)$) \cite{Baikov:2005rw}. Additionally, the electroweak (EW) corrections have been known for some time~\cite{Dabelstein1992,Kataev:1997cq}, as well as the mixed QCD$\times$EW corrections ($\mathcal{O}(\alpha \alpha_s)$) \cite{Kataev:1997cq,Mihaila:2015lwa}\footnote{Very recently, two-loop master integrals for the mixed QCD$\times$EW corrections for the Higgs-top Yukawa coupling contributions to \hbb have also been computed~\cite{Chaubey:2019lum}.}. It is thus fair to say that the theoretical knowledge of the inclusive partial decay width for \hbb is at an advanced level, with accuracies in the desired per-mille range. 
In order to study the Higgs in a collider setting it is also desirable to have theoretically- precise differential predictions, which allow for the application of experimental phase-space cuts for arbitrary infrared-safe observables. In this case our knowledge is not as advanced as at the inclusive level. Fully-differential predictions at NNLO in QCD were computed several years ago \cite{Anastasiou:2011qx,DelDuca:2015zqa,Bernreuther:2018ynm}, while more recent studies~\cite{Ferrera:2017zex,Caola:2017xuq} have focused on interfacing the decay at this order to $VH$ production, which is also known at NNLO in QCD~\cite{Ferrera:2011bk,Ferrera:2014lca,Campbell:2016jau}. The principal aim of this paper is to extend the knowledge of the \hbb decay differentially to \nnnlo accuracy.

Significant progress has been made over the past five years in regards to the computation of differential predictions at \nnlo accuracy in QCD. For most $2\rightarrow 2$ LHC processes 
NNLO predictions have been computed, and currently the frontier lies in the computation of the challenging $2\rightarrow 3$ two-loop corrections. 
A crucial aspect of this advancement has come from an increased ability to deal with the infrared (IR) divergences which affect the component parts of a NNLO calculation (but cancel
upon summation in an IR-safe observable). A novel way of dealing with IR divergences at NNLO was presented in Ref.~\cite{Cacciari:2015jma} and is now known as the Projection-to-Born (P2B) method. This method, initially applied to vector boson fusion (VBF), uses the knowledge of the inclusive cross section of the process under consideration and of the exclusive cross section of the process with one extra final-state jet to construct local counter-terms for the matrix elements, projected onto a LO phase space. At NNLO this method has 
since been applied to VBF production of two Higgs bosons \cite{Dreyer:2018rfu}. An alternate approach to pursuing NNLO calculations is to utilize physical observables and factorization theorems to construct non-local counter-terms. 
One such approach, known as $N$-jettiness slicing~\cite{Boughezal:2015dva,Gaunt:2015pea}, uses the $N$-jettiness~\cite{Stewart:2010tn} variable together with a factorization theorem derived from Soft Collinear Effective Field Theory (SCET)~\cite{Bauer:2000ew,Bauer:2000yr,Bauer:2001ct,Bauer:2001yt} to perform NNLO 
calculations.

Compared to NNLO, very few processes are known differentially at \nnnlo accuracy, although significant progress has been made over the last year. One of the flagship LHC processes, Higgs production, has recently been computed differentially at this order \cite{Cieri:2018oms} (using a non-local $q_T$-based subtraction method~\cite{Catani:2007vq}) and analytic results for the pseudo-rapidity distribution have also been computed \cite{Dulat:2018bfes,Dulat:2017prg}. These results are built upon our knowledge of the inclusive Higgs-production cross section at this order~\cite{Anastasiou:2015ema,Anastasiou:2016cez}. The P2B method has also been deployed at N$^3$LO, specifically for jet production in deep inelastic scattering \cite{Currie:2018fgr,Gehrmann:2018odt} and, for certain differential distributions, VBF and VBF di-Higgs \cite{Dreyer:2016oyx,Dreyer:2018qbw}.

The aim of this paper is to provide, for the first time, fully-differential predictions for the \hbb decay at \nnnlo accuracy. Herein we focus on the contributions with the most challenging infrared structure, namely those that are proportional to $y_b^2$. We will deploy the P2B method mentioned above and present a first application of this method in conjunction with a non-local subtraction mechanism ($N$-jettiness slicing in our case) at both \nnlo and N$^3$LO. Our paper is constructed as follows. In Section \ref{sec:calO} we present a discussion of the general framework for our calculation. We detail the P2B+SCET method in Section \ref{sec:IR} and first validate our results using the \hbb process at \nnlo. We use our calculation to make predictions for a variety of physical observables at \nnnlo accuracy in Section \ref{sec:results} and draw our conclusions in Section \ref{sec:conc}.

\section{Overview of the calculation}
\label{sec:calO}

\begin{figure}
\begin{center}
\includegraphics[width=15cm]{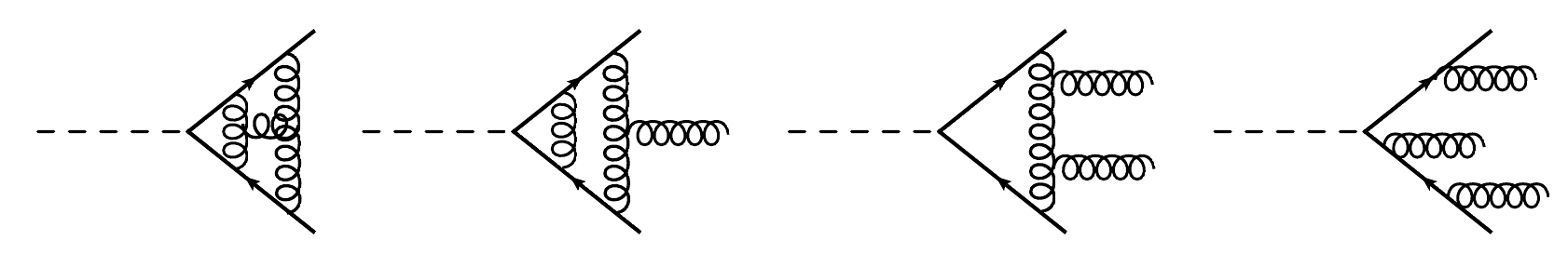}
\caption{Representative Feynman diagrams that enter our calculation of \hbb at $\mathcal{O}(\alpha_s^3)$ accuracy.}
\label{fig:FeynDH3}
\end{center}
\end{figure}

A general overview of our theoretical setup is included in our companion paper on the calculation of $H\rightarrow b\overline{b}j$ at NNLO accuracy \cite{Mondini:2019vub}. Here we provide a short summary for completeness. 
Representative Feynman diagrams included in our calculation of \hbb at \nnnlo are shown in Fig.~\ref{fig:FeynDH3}. At this order there are four phase-space configurations that contribute. The two-body phase space includes terms of up to three loops (which have been computed in Ref.~\cite{Gehrmann:2014vha}), while the remaining phase spaces correspond to those with three or more partons in the final state and are the component pieces needed for the calculation of \hbbj at NNLO. 
In our calculation we will set the $b$-quark mass to zero kinematically, but retain it in the Yukawa coupling. A comparison of the radiative corrections at NLO with or without the $b$-mass phase-space effects was first
performed nearly forty years ago \cite{Braaten:1980yq}. It was shown that the sizable differences between the full and ``massless'' theories arising from the $b$-mass terms can be compensated by running the $b$-mass to the Higgs scale (and thus recapturing some of the missing logarithms of the form $\log{(m^2_b/m^2_H)}$). Dropping the $b$-quark mass kinematically results in dramatic simplifications in the calculation of the inclusive partial width, which in the case of \hbb is known up to $\mathcal{O}(\alpha_s^4)$ in the massless theory. 
 
In this work our primary interest lies in computing the \hbb process differentially at N$^3$LO. At this order, the partial width can be written as follows: 
\begin{align}
\Gamma^{{\text{N3LO}}}_{H\rightarrow b\overline{b}} &= y_b^2 A_{b} + {\alpha_s} y_b^2 B_{b} + \alpha_s^2 \left(y_b^2 C_{b}+y_b y_t C_{bt} \right) \notag \\ &\quad +\alpha_s^3 \left(y_b^2 D_{b} +y_b y_t D_{bt} + y_t^2 D_{t} \right) + \mathcal{O}(\alpha_s^4) \, ,
\label{eq:gamman3}
\end{align}
where we have explicitly expanded in terms of both $\alpha_s$ and the Yukawa couplings to the bottom and top quark $y_b$ and $y_t$ respectively. The dependence on the top-quark mass first comes in at NNLO and corresponds to diagrams in which the Higgs boson couples to a closed loop of top quarks. These diagrams can then interfere with the LO diagram to create a mixed $y_b y_t$ term at $\mathcal{O}(\alpha_s^2)$. In our theoretical framework this interference is exactly zero due to the requirement of a helicity flip between the massless bottom quarks (since the bottom quarks couple to a spin-1 gluon in the $y_t$ term and to the scalar Higgs in the $y_b$ term). Such an interference term mandates a mass inclusion kinematically to be non-vanishing and is therefore not present in our calculation. In other words, the interference terms are suppressed by a power of $m_b/m_H$. However, since the ratio $y_t/y_b$ is large, this mixed $y_b y_t$ term is phenomenologically relevant. It is IR finite, and a commonly-used approximation is to integrate out the top-quark loop and thus work in an effective theory in which there is a clear hierarchy of scales $m_b << m_H << m_t$ \cite{Surguladze:1994gc,Chetyrkin:1995pd}. In this approximation the mixed term accounts for around 30\% of the \nnlo correction. Given that 
$m_H$ is not dramatically lighter than $m_t$, one may also worry about missing terms that are formally of order $(m_H/m_t)^4$ and could therefore result in a significant correction. 
Such a study was recently 
undertaken \cite{Primo:2018zby} keeping the exact dependence on $m_b$, $m_t$, and $m_H$, and found that the difference with respect to the exact form of the NNLO partial width are indeed small and can be neglected at the inclusive and differential level to good accuracy.  
At $\mathcal{O}(\alpha_s^3)$ a second class of diagrams enters. This contribution corresponds to diagrams in which the Higgs does not couple to the final-state $b$ quarks at all, but instead is proportional to the closed loop squared, thus creating a term proportional to $y_t^2$ at this order in Eq.~\eqref{eq:gamman3}. Additionally, the interference term which arose at NNLO now receives corrections and develops a more intricate IR structure.  
The $y_t^2$ term has particularly troublesome IR behavior since it does not factor onto the tree-level \hbb, but instead factors onto $H\rightarrow gg$. For this term there is also no helicity suppression and therefore this contribution is large and relevant for phenomenology. The Higgs coupling to partons through a top-quark loop, integrated out via an EFT approach, has been well studied in the literature \cite{Chen:2014gva,Boughezal:2015aha,Boughezal:2015dra} and is not the principal aim of this paper (where we focus on the $y_b^2$ term which has a more complicated IR structure at N$^3$LO). However, we note that these terms should be included before a full phenomenological study at \nnnlo can be completed. We leave this work to a future study, stressing that the terms that we neglect are at most NLO (for $y_b y_t$) and therefore readily amenable using existing tools to implementation in a future Monte Carlo generator.

\section{Regulation of infrared divergences at \nnnlo}
\label{sec:IR}

In this section we discuss the methods we utilize to regulate the IR singularities present in our \nnnlo calculation. We primarily focus on the 
P2B method, since the $N$-jettiness slicing method is discussed in more detail in our companion paper \cite{Mondini:2019vub}. Firstly, we recap the 
inclusive partial width, which is a prerequisite for the P2B method we use here. 

\subsection{The inclusive partial width} 

An ingredient for our calculation is the inclusive decay width for \hbb at \nnnlo. This was originally computed over two decades ago \cite{Chetyrkin:1996sr} and is now known up to 
N$^{4}$LO accuracy \cite{Baikov:2005rw}. At $\mathcal{O}(\alpha_s^3)$ the inclusive partial width $\Gamma_{H\rightarrow b\overline{b}}$ can be written as follows
\begin{eqnarray}
\Gamma^{\rm{N3LO}}_{H\rightarrow b\overline{b}} = \Gamma^{{\rm{LO}}}_{H\rightarrow b\overline{b}} + \Delta \Gamma^{\rm{NLO}}_{H\rightarrow b\overline{b}} 
+  \Delta \Gamma^{\rm{NNLO}}_{H\rightarrow b\overline{b}} +  \Delta \Gamma^{\rm{N3LO}}_{H\rightarrow b\overline{b}} +\mathcal{O}(\alpha_s^4) \, .
\end{eqnarray}
The LO partial width is defined as
\begin{eqnarray}
\Gamma^{\rm{LO}}_{H\rightarrow b\overline{b}} = 
\frac{y_b^2  m_H N_c}{8 \pi}
\end{eqnarray}
with $y_b \equiv y_b(\mu)$ the bottom Yukawa coupling at the renormalization scale $\mu$, $m_H$ the Higgs mass, and $N_c$ the number of colors, while the corrections at each order can be written as 
\begin{eqnarray}
\Delta \Gamma^{{\rm{N}}^n{\rm{LO}}}_{H\rightarrow b\overline{b}}  = \Gamma^{{\rm{LO}}}_{H\rightarrow b\overline{b}} \left(\frac{\alpha_s}{\pi}\right)^n\Gamma^{(n)}_{H\rightarrow b\overline{b}}
\end{eqnarray}
with $\alpha_s \equiv \alpha_s(\mu)$. The coefficients $\Gamma^{(n)}_{H\rightarrow b\overline{b}}$ up to $n=3$ are: 
\begin{eqnarray}
\Gamma^{(1)}_{H\rightarrow b\overline{b}} &=&  s_1 + 2  \gamma^{0}_m L \label{eq:ga1coeff} \\
\Gamma^{(2)}_{H\rightarrow b\overline{b}} &=&  s_2 +L  \left(s_1\beta_0 + 2 s_1 \gamma^0_m + 2 \gamma^1_m\right) +
   L^2 \left(\beta_0 \gamma^0_m + 2 (\gamma^0_m)^2\right) \\
   \Gamma^{(3)}_{H\rightarrow b\overline{b}} &=& 
   s_3 + L \left(2 s_2 \beta_0 + s_1\beta_1 + 2 s_2 \gamma^0_m + 
     2 s_1 \gamma^1_m + 2 \gamma^2_m \right) \nonumber\\&&
  + L^2 \left(s_1\beta_0^2 + 
     3 s_1 \beta_0 \gamma^0_m + \beta_1 \gamma^0_m + 
     2 s_1 (\gamma^0_m)^2 + 2 \beta_0 \gamma^1_m + 
     4 \gamma^0_m\gamma^1_m\right) \nonumber\\&&+ 
L^3 \left(\frac{2}{3} \beta_0^2 \gamma^0_m + 2 \beta_0(\gamma^0_m)^2 + 
     \frac{4}{3} (\gamma^0_m)^3\right) \label{eq:ga3coeff}
     \end{eqnarray} 
where $L = \log{(\mu^2/m_H^2)}$ and the explicit expressions for $s_i$, $\beta_i$ and $\gamma^i_m$ are presented in Appendix~\ref{sec:Incw}.
For reference, at $\mu=m_H$ the inclusive partial width numerically evaluates to  
\begin{eqnarray}
\Gamma^{\rm{N3LO}}_{H\rightarrow b\overline{b}}(\mu=m_H) = \Gamma^{LO}_{H\rightarrow b\overline{b}}\left[1+ 5.66667 \left(\frac{\alpha_s}{\pi}\right)  +  29.1467 \left(\frac{\alpha_s}{\pi}\right)^2 +  41.7576 \left(\frac{\alpha_s}{\pi}\right)^3\right] .
\end{eqnarray}
Finally, we will employ the following definition of the \nnnlo coefficient for the inclusive width, which reinstates the dependence on the LO phase space (evaluated in $d=4$ dimensions):
\begin{eqnarray}
\Delta \Gamma^{\rm{N3LO}}_{H\rightarrow b\overline{b}} &=& \left(\frac{\alpha_s}{\pi}\right)^3 \int 8\pi \, \Gamma^{\rm{LO}}_{H\rightarrow b\overline{b}} \Gamma^{(3)}_{H\rightarrow b\overline{b}} \,d \Phi_2 \\
                                      & = & \int \Delta\hat{\Gamma}^{\rm{N3LO}}_{H\rightarrow b\overline{b}} \, d \Phi_2 \, . \label{eq:hatdeltan3lo}
\end{eqnarray}

\subsection{Projection to Born at \nnnlo}

The \hbb differential decay width at \nnnlo is constructed as follows
\begin{eqnarray}
\frac{d \,\Delta\Gamma^{{\rm{N3LO}}}_{H\rightarrow b\overline{b}}}{d\,\mathcal{O}_m} &= &\int d\Gamma^{VVV}_{H\rightarrow b\overline{b}} F^{m}_{2}(\Phi_2)  d\Phi_2 +  \int d\Gamma^{RVV}_{H\rightarrow b\overline{b}} F^{m}_{3}(\Phi_3)  d\Phi_3 
 \nonumber\\ &&+ \int d\Gamma^{RRV}_{H\rightarrow b\overline{b}} F^{m}_{4}(\Phi_4)  d\Phi_4 +  \int d\Gamma^{RRR}_{H\rightarrow b\overline{b}} F^{m}_{5}(\Phi_5)  d\Phi_5 \, ,
 \label{eq:pwn3lo}
\end{eqnarray}
where $d\Gamma^{VVV}_{H\rightarrow b\overline{b}}$ represents the triple-virtual contribution to the decay width, $d\Gamma^{RVV}_{H\rightarrow b\overline{b}}$ the real double-virtual contribution, $d\Gamma^{RRV}_{H\rightarrow b\overline{b}}$ the double-real virtual contribution, and $d\Gamma^{RRR}_{H\rightarrow b\overline{b}}$ the triple-real contribution. Each parton-level contribution belongs to a different phase space $\Phi_i$ (with $i=2,\dots,5$ respectively) over which it is integrated. The measurement function $F^{m}_{i}(\Phi_i)$ uses an IR-safe jet algorithm to cluster the $i$ final-state partons onto $m$ final-state jets and thus defines the observable $\mathcal{O}_m$.  The triple-virtual contribution contains explicit poles in the dimensional regularization parameter $\epsilon=(4-d)/2$ (with $d$ the number of space-time dimensions), whereas the triple-real term contains only implicit poles that become manifest as 
at least one and at most three particles become unresolved. The RVV and RRV contributions consist of mixtures of explicit $\epsilon$ poles and implicit phase-space singularities. 
The triple-virtual piece can be obtained from the results presented in Ref.~\cite{Gehrmann:2014vha}, and real double-virtual in Refs.~\cite{Ahmed:2014pka,Mondini:2019vub}, while the calculation of $H\rightarrow b\overline{b} j$ at NNLO accuracy is discussed in our companion paper \cite{Mondini:2019vub}. This means that all the individual terms in Eq.~\eqref{eq:pwn3lo} are known, but need IR regulation to be combined in a physically-meaningful way. 

We define the Born-projected inclusive partial width as follows, 
\begin{eqnarray}
\frac{d\Delta \Gamma^{{\rm{N3LO,\,inc}}}_{H\rightarrow b\overline{b}}}{d \mathcal{O}^B_m}  = \int  \Delta\hat{\Gamma}^{\rm{N3LO}}_{H\rightarrow b\overline{b}} F_2^{m}(\Phi_B) d \Phi_B 
\end{eqnarray}
where $\Phi_B = \Phi_2$ corresponds to the LO phase space and $\mathcal{O}^B_m$ represents the observable $\mathcal{O}_m$ evaluated for LO kinematics. We note the insertion of the two-body measurement function $F_2^{m}(\Phi_B)$ into the integrand in relation to Eq.~\eqref{eq:hatdeltan3lo}. Expanding out the various component pieces of the Born-projected inclusive width yields the following 
\begin{eqnarray}
\frac{d\Delta \Gamma^{{\rm{N3LO,\,inc}}}_{H\rightarrow b\overline{b}}}{d \mathcal{O}^B_m}   &= &\int d\Gamma^{VVV}_{H\rightarrow b\overline{b}} F^{m}_{2}(\Phi_B)  d\Phi_2 +  \int d\Gamma^{RVV}_{H\rightarrow b\overline{b}} F^{m}_{2}(\Phi_B) d\Phi_3 
 \nonumber\\ &&+ \int d\Gamma^{RRV}_{H\rightarrow b\overline{b}}   F^{m}_{2}(\Phi_B) d\Phi_4 +  \int d\Gamma^{RRR}_{H\rightarrow b\overline{b}} F^{m}_{2}(\Phi_B)  d\Phi_5 \, .
 \label{eq:pwn3lo_inc}
\end{eqnarray}
The fully-differential \nnnlo coefficient can then be written as
\begin{eqnarray}
\frac{d \,\Delta\Gamma^{{\rm{N3LO}}}_{H\rightarrow b\overline{b}}}{d\,\mathcal{O}_m}  = \frac{d \,\Delta\Gamma^{{\rm{N3LO,\,inc}}}_{H\rightarrow b\overline{b}}}{d\,\mathcal{O}^B_m}  - \frac{d \,\Delta\Gamma^{{\rm{NNLO }}}_{H\rightarrow b\overline{b}j}}{d\,\mathcal{O}^B_m} + \frac{d \,\Delta\Gamma^{{\rm{NNLO}}}_{H\rightarrow b\overline{b}j}}{d\,\mathcal{O}_m} \label{eq:masterp2bn3lo}
\end{eqnarray}
where explicitly
\begin{align}
\frac{d \,\Delta\Gamma^{{\rm{NNLO }}}_{H\rightarrow b\overline{b}j}}{d\,\mathcal{O}_m} &= \int d\Gamma^{RVV}_{H\rightarrow b\overline{b}} F^{m}_{3}(\Phi_3) d\Phi_3 + \int d\Gamma^{RRV}_{H\rightarrow b\overline{b}} F^{m}_{4}(\Phi_4) d\Phi_4 \notag \\ &\quad + \int d\Gamma^{RRR}_{H\rightarrow b\overline{b}} F^{m}_{5}(\Phi_5) d\Phi_5 \label{eq:nnlo3jdiff}
\end{align}
and
\begin{align}
\frac{d \,\Delta\Gamma^{{\rm{NNLO }}}_{H\rightarrow b\overline{b}j}}{d\,\mathcal{O}^B_m} &= \int d\Gamma^{RVV}_{H\rightarrow b\overline{b}} F^{m}_{2}(\Phi_B) d\Phi_3 + \int d\Gamma^{RRV}_{H\rightarrow b\overline{b}} F^{m}_{2}(\Phi_B) d\Phi_4 \notag \\ &\quad + \int d\Gamma^{RRR}_{H\rightarrow b\overline{b}} F^{m}_{2}(\Phi_B) d\Phi_5 \, . \label{eq:nnlo3jdiffB}
\end{align}
Eq.~\eqref{eq:masterp2bn3lo} represents the master equation for the Projection-to-Born technique \cite{Cacciari:2015jma,Currie:2018fgr} and is equivalent to Eq.~\eqref{eq:pwn3lo} by explicitly substituting Eqs.~\eqref{eq:pwn3lo_inc}, \eqref{eq:nnlo3jdiff}, and \eqref{eq:nnlo3jdiffB}. It can finally be rearranged as follows
\begin{eqnarray}
\frac{d \,\Delta\Gamma^{{\rm{N3LO}}}_{H\rightarrow b\overline{b}}}{d\,\mathcal{O}_m}    &= &\int  \Delta\hat{\Gamma}^{\rm{N3LO}}_{H\rightarrow b\overline{b}} F_2^{m}(\Phi_B) d \Phi_B \nonumber\\
&&
+  \int d\Gamma^{RVV}_{H\rightarrow b\overline{b}} \,[ F^{m}_{3}(\Phi_3)-F^{m}_{2}(\Phi_B) ] \, d\Phi_3 
 \nonumber\\ &&
 +\int d\Gamma^{RRV}_{H\rightarrow b\overline{b}} \,[ F^{m}_{4}(\Phi_4) - F^{m}_{2}(\Phi_B) ] \, d\Phi_4  \nonumber\\&&
 +  \int d\Gamma^{RRR}_{H\rightarrow b\overline{b}} \,[ F^{m}_{5}(\Phi_5) - F^{m}_{2}(\Phi_B) ] \, d\Phi_5 \, .
 \label{eq:p2bmast}
\end{eqnarray}
Inspection of the above formula reveals that the P2B subtraction regularizes singularities which cancel when an implicit pole turns to an explicit one via phase-space integration, i.e.~this subtraction accounts for the ``last emission''. 
Based on the above equation, the full \nnnlo \hbb coefficient can be readily computed provided that the NNLO $H\rightarrow b\overline{b} j$ differential partial width is available in a suitable format. More specifically, since the P2B method above regulates the singularities associated with the last emission, all the other IR divergences present in the last three lines of Eq.~\eqref{eq:p2bmast} (namely in the construction of the differential cross section of the process with one extra final-state jet) have to be previously regulated and canceled by means of a different subtraction scheme. Thus far, applications of the P2B method have utilized Catani-Seymour dipoles \cite{Catani:1996vz} (for applications at NNLO) and antenna subtraction \cite{GehrmannDeRidder:2005cm} (for applications at N$^3$LO) for this purpose. Both these regulators are clearly a good fit for the method, since neither explicitly requires a jet in the construction of the local counter-terms. Thus far no method that employs a jet-based physical observable to regulate divergences at NNLO has been applied to P2B. We address this in the subsequent section. 

\subsection{P2B with $N$-jettiness slicing} 

At first inspection the application of Eq.~\eqref{eq:p2bmast} with $N$-jettiness slicing seems problematic, since the application of $N$-jettiness slicing requires the definition of a jet observable (in this case 3-jettiness) in order to operate. Here we address this issue, starting with a brief summary of the method which is by now well established for NNLO calculations. 

The central idea of any slicing-based method is to consider an observable which allows one to separate the computation into two parts. At NNLO, the first part will contain all of the doubly-unresolved regions of the phase space and will be computed using a simplifying approximation (typically a factorization theorem). The second region will capture all of the singly-unresolved and fully-resolved regions of phase space and thus corresponds to a NLO calculation with one additional parton in the final state. In $N$-jettiness slicing, the separating variable is the $N$-jettiness variable $\tau_N$ \cite{Stewart:2010tn}. For an $n$-parton event it is defined as 
\begin{eqnarray}
\tau_N = \sum_{j=1,n} \min_{i=1,2,N} \left\{ \frac{2 q_i \cdot p_{j}}{Q_i}\right\}
\label{eq:tauN}
\end{eqnarray}
where $p_j$ represent the momenta of the $n$ partons, while $q_i$ represent the momenta of the $N$ most energetic jets clustered with any IR-safe jet algorithm (in our case the Durham jet algorithm~\cite{Brown:1990nm,Catani:1991hj}). $Q_i$ are the hard scales in the process, which we take as $Q_i = 2E_i$ with $E_i$ the energy of the $i$-th jet. In order to separate the phase space into two regions, we introduce a variable $\tau^{\rm{cut}}_N$. In the region $\tau_N > \tau^{\rm{cut}}_N$ at least one of the $n$ partons is resolved (so that the term $2 q_i \cdot p_{j}$ in Eq.~\eqref{eq:tauN} is non-vanishing). The NNLO decay width for a generic $H\rightarrow N j$ process can be then computed in this region as the NLO calculation of the $H\rightarrow (N+1) j$ process. On the other hand, in the region $\tau_N < \tau^{\rm{cut}}_N$ no parton is resolved and the NNLO decay width can be approximated with the following convolution, derived from SCET \cite{Stewart:2010tn,Stewart:2009yx}: 
\begin{eqnarray}
\Gamma^{\rm{NNLO}}_{H\rightarrow N j} \left( \tau_N < \tau^{\rm{cut}}_N \right) \approx \int \prod_{i=1}^{N} \mathcal{J}_i \otimes \mathcal{S} \otimes \mathcal{H} + \mathcal{O}(\tau^{\rm{cut}}_N) \, .
\label{eq:bewtau}
\end{eqnarray}
In the above equation the terms $\mathcal{J}_i$ represent the jet functions \cite{Becher:2006qw,Becher:2010pd}, $\mathcal{S}$ denotes the soft function for $N$ colored partons, and $\mathcal{H}$ is the process-specific hard function. In our application of $N$-jettiness slicing we consider $N=3$ and therefore we need the NNLO 1-jettiness soft function with arbitrary kinematics \cite{Campbell:2017hsw}\footnote{See also Refs.~\cite{Gaunt:2015pea,Boughezal:2015eha}.} and the hard function computed in our companion paper \cite{Mondini:2019vub}. We also note that Eq.~\eqref{eq:bewtau} is accurate up to terms of $\mathcal{O}(\tau^{\rm{cut}}_N)$, which formally vanish in the limit $\tau^{\rm{cut}}_N \rightarrow 0$. One should therefore set $\tau^{\rm{cut}}_N$ as small as possible to ensure the validity of the factorization formula.

In order to apply $N$-jettiness slicing in conjunction with Eq.~\eqref{eq:p2bmast}, let us consider the types of partonic configurations that can occur in our calculation. As an example, let us focus on the five-parton phase space (the triple-real contribution in Eq.~\eqref{eq:p2bmast}). In the Higgs rest frame, after jet clustering each phase-space event will belong to one of four possible topologies: a two-, three-, four-, or five-jet topology. We assume now that we are calculating an observable that requires the complete N$^3$LO technology and thus we fix the measurement function to demand exactly $m=2$ jets (any observable with three or more jets requires at most a NNLO calculation). In the triple-real contribution to Eq.~\eqref{eq:p2bmast} there are two measurement functions: $F_5^2(\Phi_5)$ and $F_2^2(\Phi_B)$. The latter will always produce two jets (in the rest frame) since it acts on the LO phase space $\Phi_B$. It is therefore unaffected by the number of jets obtained upon  clustering of the five-parton phase space (assuming for now that no $p_T$ or rapidity cuts are applied to the LO phase space). On the other hand, $F_5^2(\Phi_5)$ will pick out the various jet topologies given an input jet algorithm, in this case vetoing any event with more than two jets (since we fixed $m=2$). This means that upon generation of a phase-space event there are two possibilities: $a)$ the five-parton event corresponds to a $\ge 3$-jet topology, is vetoed by $F_5^2(\Phi_5)$ and therefore only the P2B subtraction term is non-zero, or $b$) the parton-level event produces two jets. In the latter case both terms in the last line of Eq.~\eqref{eq:p2bmast} survive, producing events with exactly-opposite weights, with the measurement functions applied on different phase spaces (which match in the triple-unresolved limit producing the desired subtraction). 

For events belonging to category $a)$ it is straightforward to compute the 3-jettiness variable $\tau_3$ and apply the cut $\tau^{\rm{cut}}_3$ since there are (at least) three jets in the event (this is indeed simply a rephrasing of the existing NNLO methodology). Attention must be given to category $b)$ two-jet events for which it is in principle unclear how a 3-jettiness cut can be constructed. In other words, in this case we must extract a three-jet observable from events with a two-jet topology. 
In order to achieve this, we first decluster the jets (in a similar spirit to the ideas behind jet-substructure techniques). Specifically, we reverse the last stage of the clustering algorithm, resulting in exactly three sub-jets. We then apply ``$N$-subjettiness'' slicing, taking the momenta of the three sub-jets as the momenta $q_i$ in Eq.~\eqref{eq:tauN}. Crucial to the success of this approach is the lack of explicit dependence on the jet algorithm in the factorization formula of Eq.~\eqref{eq:bewtau}. Furthermore, since events in category $b)$ have zero weight as explained above, the total two-jet rate at \nnnlo inherits the overall $\tau^{\rm{cut}}_N$-dependence of the parent NNLO calculation.  In this regard, we do not expect significant worsening of the power corrections when applied to our \nnnlo calculation relative to our NNLO application. We investigate this behavior more carefully in the next section. Finally, the same line of reasoning can be applied to the double-real virtual and double-virtual real contributions. 
 
We conclude this section by defining the Born phase-space events that enter the P2B subtraction terms. For each event we simply define the following Born phase-space point:
 \begin{eqnarray}
 \Phi_B  = \{ p_1, p_2\}, \quad p_1  = \frac{m_H}{2}(1,{\bold{n}}_{j}), \quad p_2 = \frac{m_H}{2} (1,-{\bold{n}}_{j})
\end{eqnarray}
where ${\bf{n}}_{j}$ is the three-dimensional unit vector pointing in the direction of the leading jet (defined as the jet with the largest energy component). 

\subsection{Validation at NNLO}
\begin{figure}
\begin{center}
\includegraphics[width=10cm]{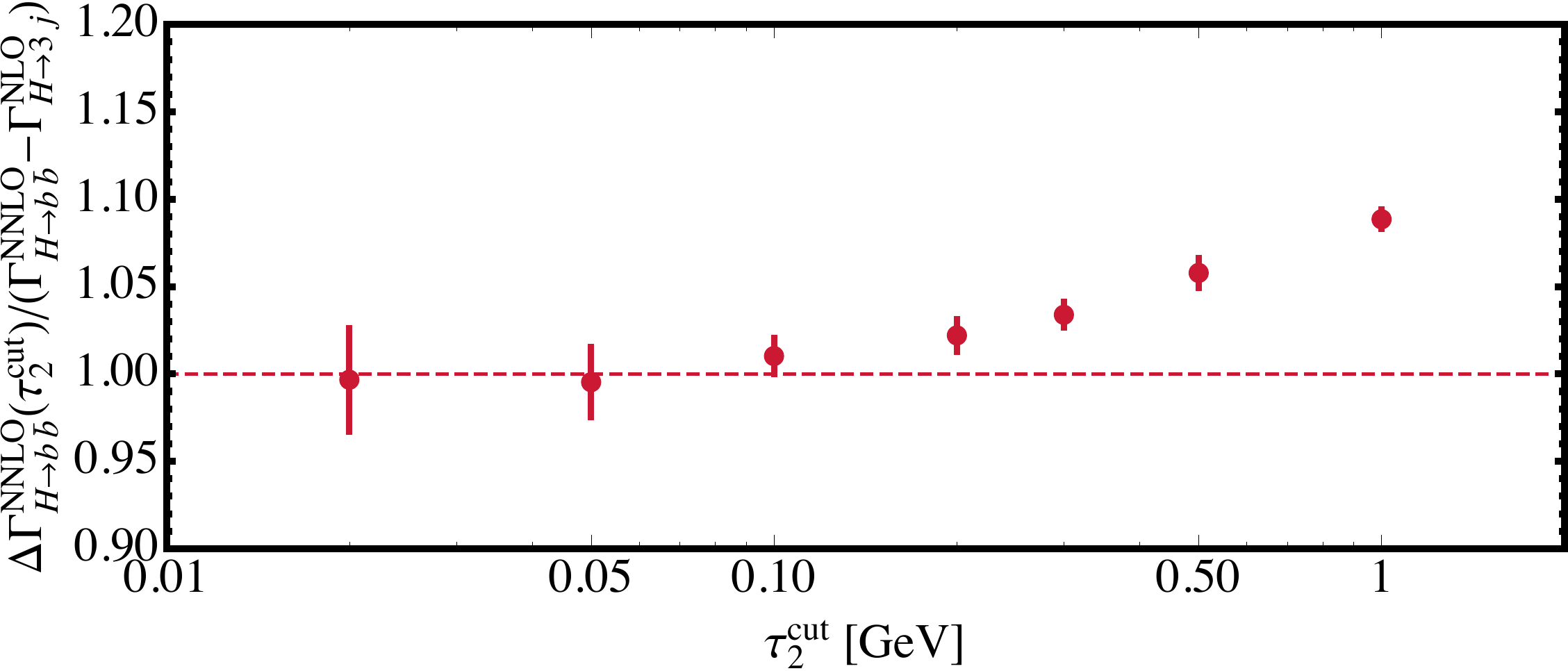}
\caption{The dependence of the \hbb \nnlo coefficient for the two-jet partial width on the $N-$jettiness slicing parameter $\tau^{\rm{cut}}_2$. The physical jet cut is set to $y_{\rm{cut}} =0.1$. The 
coefficient is normalized to the prediction obtained from the difference of the inclusive result and the \nlo (inclusive) three-jet rate.}
\label{fig:taudep_xs}
\end{center}
\end{figure}

In order to validate our implementation of the P2B method at \nnlo we have implemented an independent calculation at this order using the $N$-jettiness slicing approach. As discussed in previous sections, this method uses the predictions of SCET to establish a factorization theorem which can be used at small values of the physical $N$-jettiness observable $\tau_N$ (which in this instance corresponds to a 2-jettiness cut, $\tau_2$).  One therefore must ensure that the $\tau^{\rm{cut}}_2$ variable is taken to small enough values that the missing power corrections in Eq.~\eqref{eq:bewtau} are negligible. Our parameter choices are as follows. We take the mass of the Higgs boson to be $m_H = 125$ GeV. As input we take the mass of the $b$-quark to be $m_b =4.7$ GeV, which enters into the Yukawa coupling $y_b$ (and is set to zero kinematically). In order to compensate for higher-order effects arising from the $b$-quark mass we run the mass to the Higgs scale. At \nnlo we use the three-loop running, resulting in an effective $b$-quark mass of $m_b(m_H) = 2.94$ GeV.  Our remaining electroweak inputs are $G_F = 0.116639 \times 10^{-4}$ GeV$^{-2}$ and $m_W = 80.385$ GeV. We take $\alpha_s(m_Z) = 0.118$ and we evolve the coupling using three-loop running. For our subsequent predictions at \nnnlo we keep the three-loop running of $\alpha_s$ and $m_b$ for simplicity (the difference between three-loop and four-loop running is very small \cite{Chetyrkin:2000yt}). All of the results for partial widths in this paper are in units of MeV. Our results presented herein have been produced using a fully-flexible Monte Carlo code, for which we have extensively used the existing structure of MCFM 8.0 where applicable (specifically for phase-space generation, Catani-Seymour dipoles~\cite{Campbell:1999ah}, $N$-jettiness slicing~\cite{Boughezal:2016wmq}, and OMP and MPI compatibility~\cite{Campbell:2015qma}). Our subsequent extended Monte Carlo is thus in a suitable format to be interfaced with MCFM and be released publicly in the future. 

 As a first check on the correctness of our results we compute the \nnlo coefficient for the two-jet rate for jets clustered with the Durham algorithm \cite{Brown:1990nm,Catani:1991hj} with $y_{\rm{cut}} = 0.1$. This algorithm starts from a parton-level phase-space point and computes the following quantity ${y}_{ij}$ for all pairs of objects $i$ and $j$:
\begin{eqnarray}
y_{ij} = \frac{2 \,{\rm{min}}(E_i^2,E_j^2)(1-\cos{\theta_{ij}})}{Q^2} \, ,
\end{eqnarray}  
where $E_i$ is the energy of particle $i$, $\theta_{ij}$ is the angle between particles $i$ and $j$, and $Q$ is the hard scale of the process, which in our case is $Q=m_H$. If $y_{ij} < y_{\rm{cut}}$, the two objects are combined into a new one with four momentum $p^{\mu}_{i}+p^{\mu}_j$. The procedure is then iterated until no more clustering is possible and the final objects are classified as jets. 
\begin{figure}
\begin{center}
\includegraphics[width=10cm]{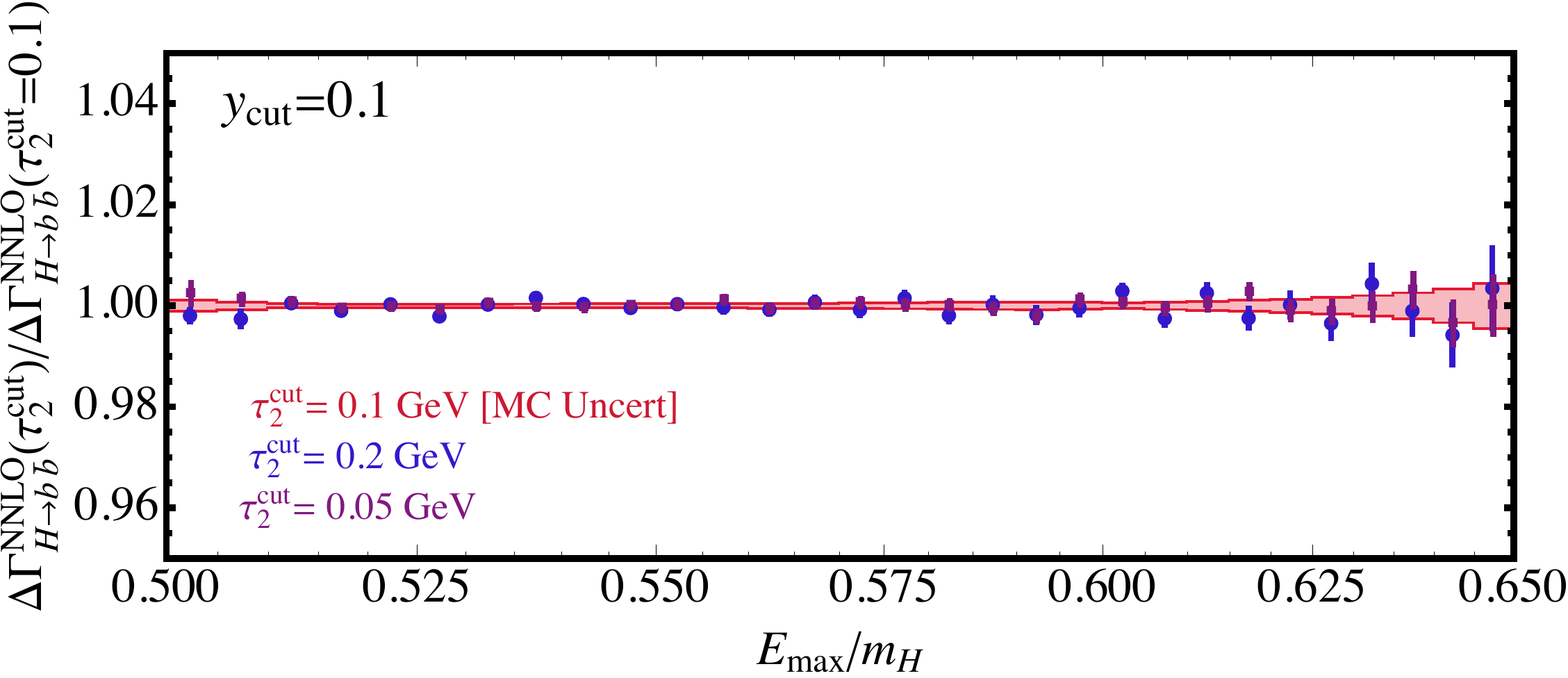}
\caption{The dependence of the differential distribution for the maximum jet energy in  the \nnlo two-jet rate on the $N-$jettiness slicing parameter $\tau^{\rm{cut}}_2$. The physical jet cut is set to $y_{\rm{cut}} =0.1$.}
\label{fig:taudep_Emax}
\end{center}
\end{figure}
In addition to the independence on the slicing parameter, a further check of our implementation of the $N$-jettiness slicing calculation of the NNLO two-jet rate can be constructed by taking the difference between the \nnlo  total inclusive rate and the inclusive three-jet rate at NLO. We compare this prediction to our results obtained with $N$-jettiness slicing in Fig.~\ref{fig:taudep_xs} observing excellent agreement 
in the asymptotic region $\tau^{\rm{cut}}_2 < 0.1$ GeV. In order to ensure that the dependence on $\tau^{\rm{cut}}_2$ in the differential distributions is also small we present the differential ratio for two different choices of $\tau^{\rm{cut}}_2$ for the $E_{\rm{max}}/m_H$ observable in Fig.~\ref{fig:taudep_Emax}. Again, we observe excellent agreement for different choices of $\tau^{\rm{cut}}_2$.  We use the prediction with $\tau^{\rm{cut}}_2 = {0.05}$ GeV for our subsequent comparisons with the P2B method. 

We now compare the predictions from $N$-jettiness slicing to our implementation of P2B at NNLO. We have implemented the P2B method at NNLO using two different subtraction methods for the NLO part of the calculation: one with Catani-Seymour dipoles, and a second one using $N$-(sub)jettiness slicing. In the Higgs rest frame the most physically-relevant observables are delta functions at LO (for example the jet energy or the jet mass). In general, there is no special direction in momentum space with which to construct more elaborate observables. In order to fully test the cancellation of IR singularities it is most useful to construct an observable which has a non-trivial distribution at LO. In this paper we therefore introduce the following two quantities: the transverse momentum of the leading jet (the jet with highest energy) $p_{T,j}^{\rm{max}}$ and the pseudo-rapidity of the jet $|\eta_j^{\rm{max}}|$. These two jet observables are measured with respect to the ``$z$''-axis which we take to be a fictitious beam axis (i.e.~we imagine that the Higgs was formed in a $\mu^+\mu^-$ collision with an operating energy $\sqrt{s} = m_H$).

\begin{figure}
\begin{center}
\includegraphics[width=7cm]{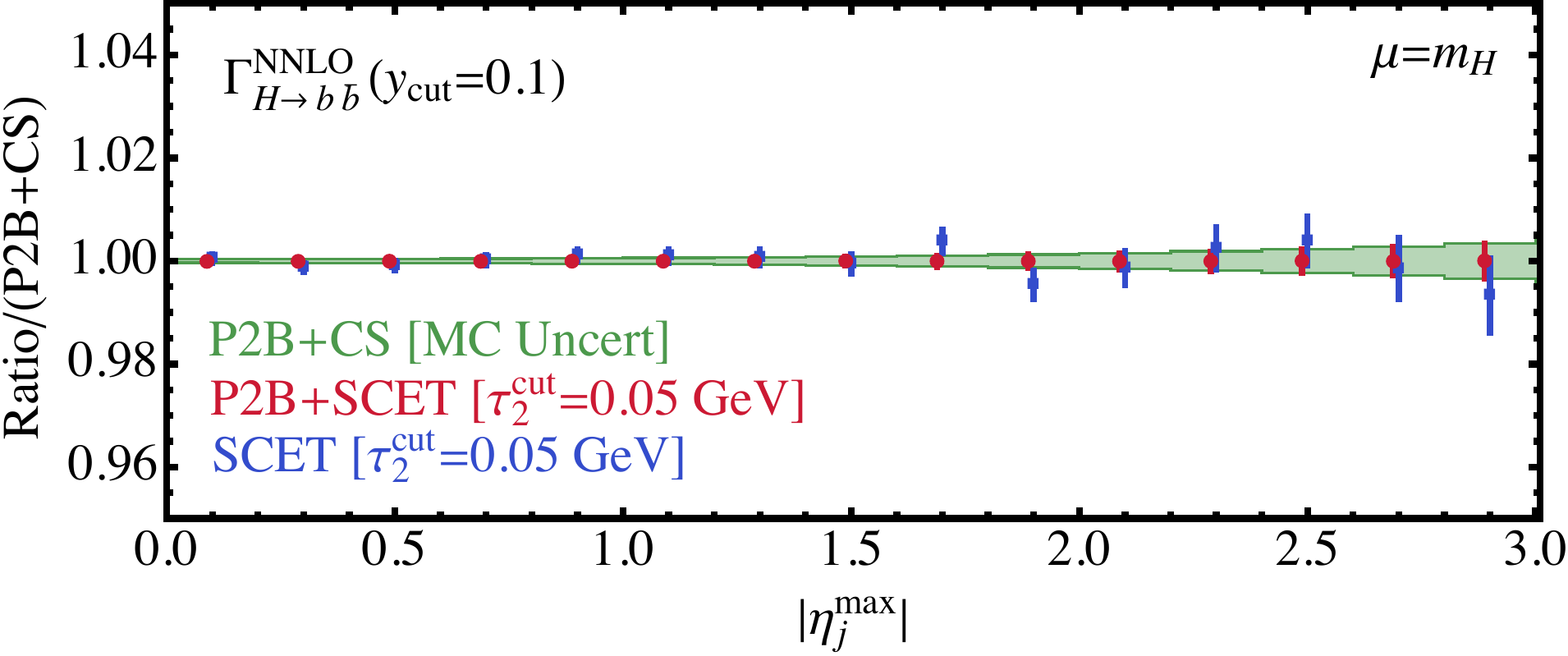}
\includegraphics[width=7cm]{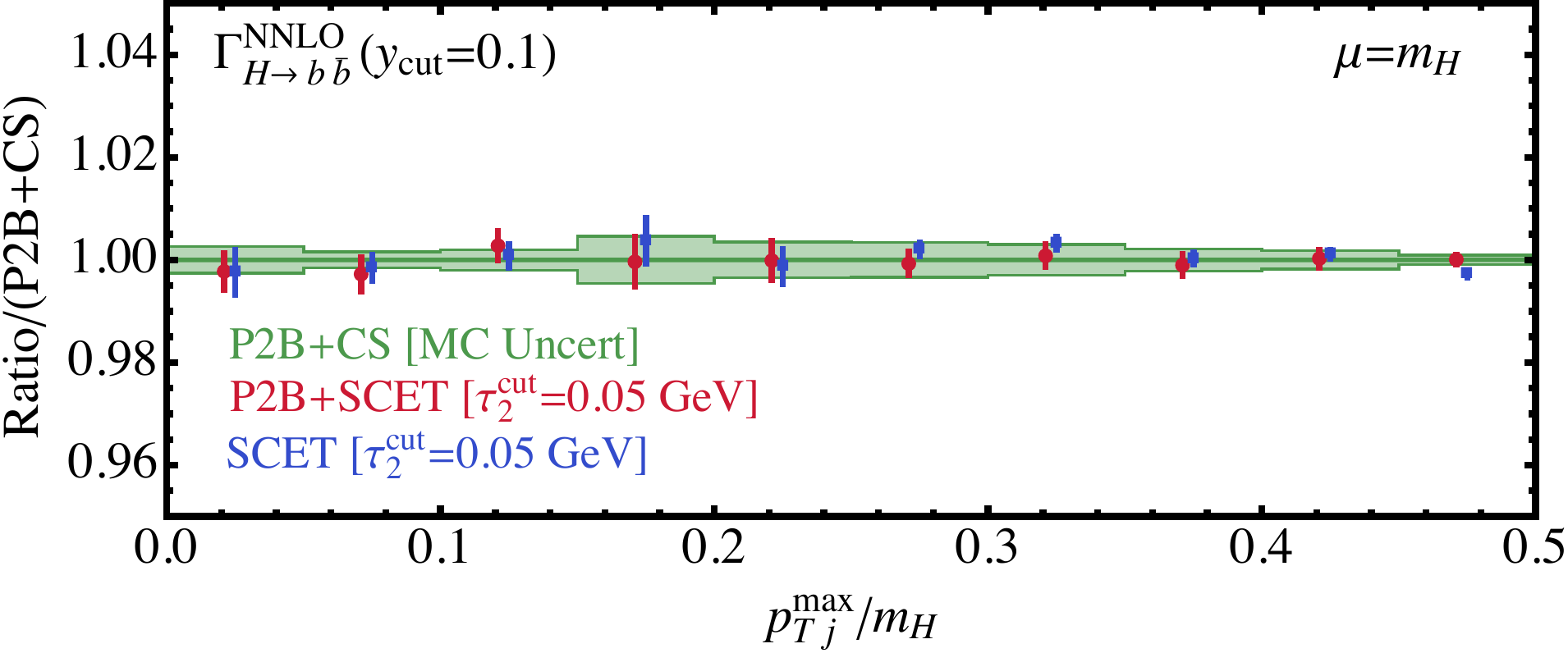}
\caption{Comparison of three different methodologies for computing the differential NNLO partial width. Shown are results obtained using Projection-to-Born with Catani-Seymour dipoles (P2B+CS), Projection-to-Born with $N$-jettiness slicing (P2B+SCET), and $N$-jettiness slicing (SCET). Results are normalized to those obtained using P2B+CS. The left-hand plot shows the pseudo-rapidity, while the right-hand plot shows the transverse momentum of the jet.}
\label{fig:P2B_val}
\end{center}
\end{figure}

The calculation of these observables at NNLO is presented in Fig.~\ref{fig:P2B_val}. We set $\mu = m_H$ for these predictions and maintain the same parameter choices as before. We choose a value of $\tau^{\rm{cut}}_2 = {0.05}$ GeV for both of the calculations which require $N$-jettiness slicing. We observe excellent agreement within the sub-percentage Monte Carlo uncertainties for all three predictions. Our proposed method of P2B+$N$-jettiness slicing is thus validated at NNLO and we proceed to use this method to obtain results at \nnnlo accuracy in the next section.

\section{Results}
\label{sec:results}

The results presented in this section are obtained using the same parameter choices as discussed in Section~\ref{sec:IR}. 
We begin by computing jet rates at $\mathcal{O}(\alpha_s^3)$. At this order, possible topologies consist of two-, three-, four-, or five-jet events, which 
are accurate respectively to N$^{3}$LO, NNLO, NLO, and LO in perturbation theory. Since the inclusive partial width is known at N$^{3}$LO, the two-jet rate can be 
inferred directly from the knowledge of the other components at their respective orders. 
Therefore, we can use the \nnlo three-jet results taken from our companion paper \cite{Mondini:2019vub}, compute the exclusive NLO four-jet and LO five-jet rates as a function of the $y_{\rm{cut}}$ parameter, and obtain the two-jet rate at N$^3$LO. 

\begin{figure}
\begin{center}
\includegraphics[width=4.5cm]{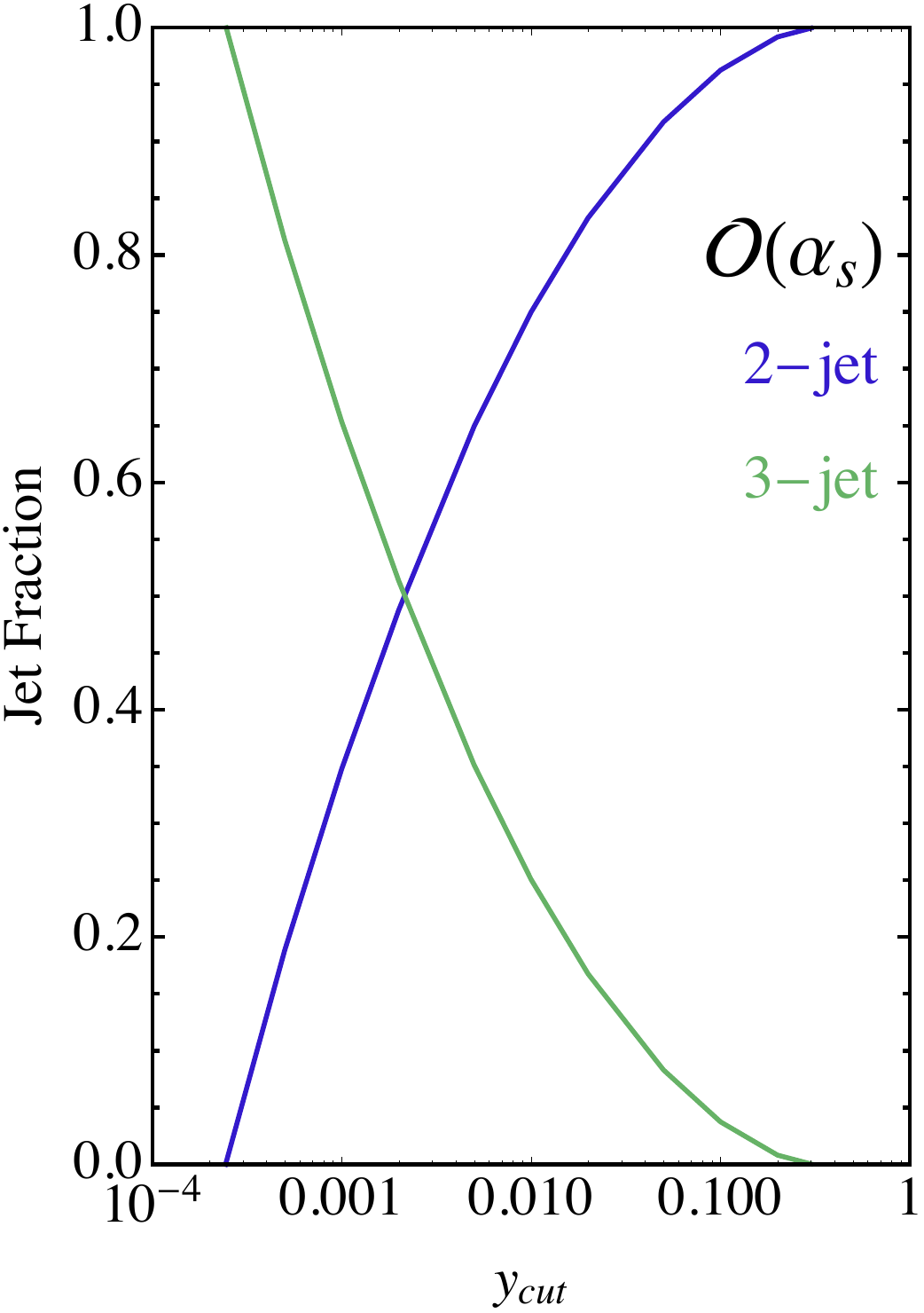}
\includegraphics[width=4.5cm]{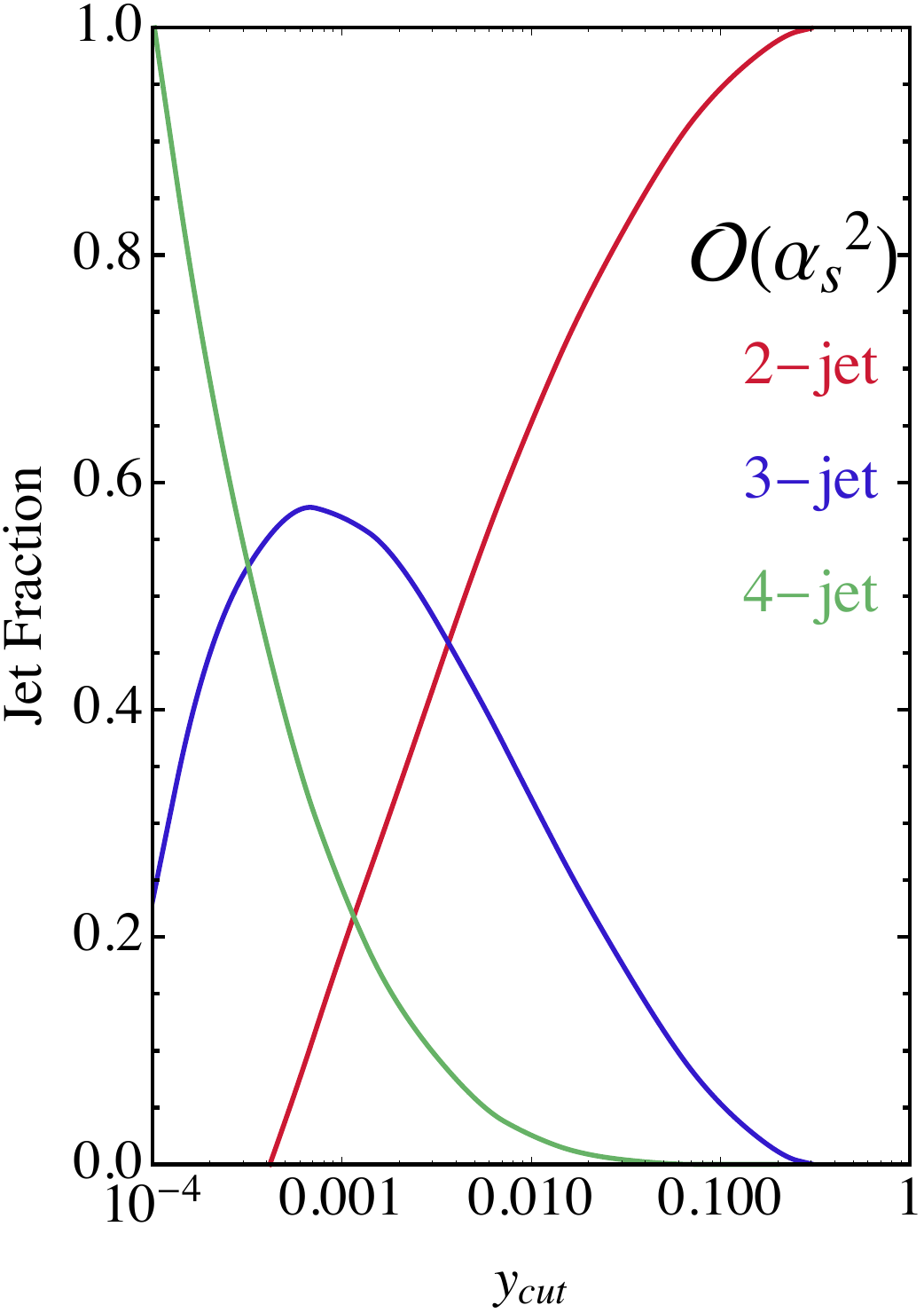}
\includegraphics[width=4.5cm]{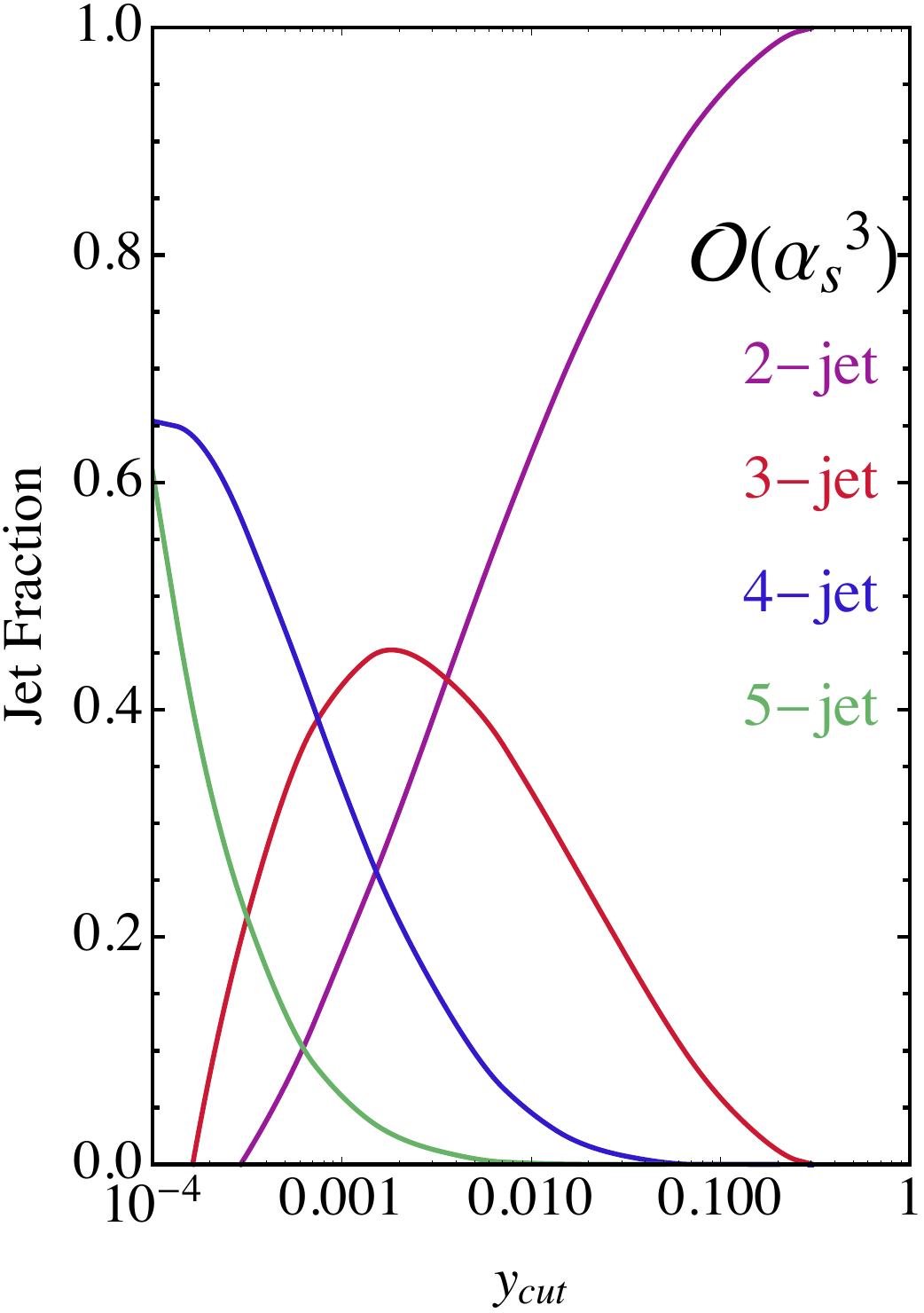}
\caption{Jet fractions at orders $\alpha_s$, $\alpha_s^2$, and $\alpha_s^3$. Each prediction is normalized to the total partial width at that order.} 
\label{fig:jetFrac}
\end{center}
\end{figure}

Our results are presented in Fig.~\ref{fig:jetFrac}, where we present the fractional jet rate at different orders in $\mathcal{O}(\alpha_s)$, each prediction being normalized to the total partial width at that order. As it may be expected, the characteristics are broadly the same as similar calculations for $e^+e^-\rightarrow Z\rightarrow $ jets computed at the same order~\cite{Weinzierl:2008iv,GehrmannDeRidder:2008ug}. For $Z\rightarrow$ jets, copious data from LEP is available for a comparison between theory and data. 
A future lepton collider should therefore be able to make the same sort of plot and compare to our predictions here. Expecting similarities with the $Z$ data, as the order in perturbation theory increases the agreement with 
data for the jet rate is expected to improve. At smaller $y_{\rm{cut}}$ the two-jet rate turns negative at each order in perturbation theory (beyond LO). However, for $\mathcal{O}(\alpha_s^3)$ the fractional rate is very small and negative for the smallest values of $y_{\rm{cut}}$ considered here. Specifically, at $y_{\rm{cut}} = 10^{-4}$ the two-jet fractional rate at NNLO is $-24\%$, whereas at \nnnlo the rate is only $-4\%$. One may therefore optimistically hope that at N$^{4}$LO the two-jet rate will remain physical to even very small values of the jet-clustering parameter.  The change in slope for small values of the jet-clustering parameter is clearly visible when comparing the NNLO plot (middle plot, red line) to the \nnnlo one (right-hand plot, purple line). 

\begin{figure}
\begin{center}
\includegraphics[width=11cm]{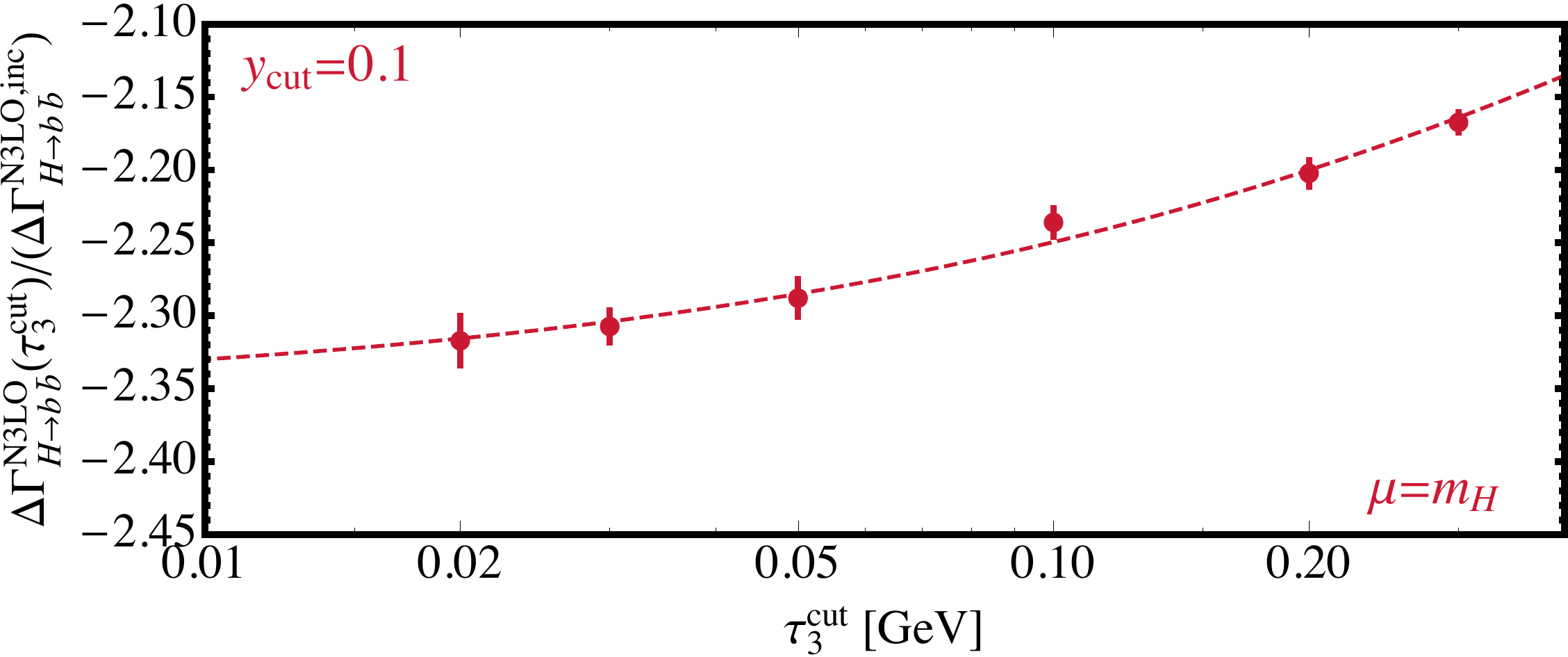}
\caption{The dependence of the \nnnlo coefficient (in units of the inclusive \nnnlo coefficient $\Delta \Gamma^{{\rm{N3LO}}}_{H\rightarrow b\overline{b}}$) on the parameter $\tau_3^{\rm{cut}}$.} 
\label{fig:taudep_n3l}
\end{center}
\end{figure}

For the remainder of this section we will turn our attention to \nnnlo predictions which cannot simply be inferred from the NNLO three-jet inclusive rate. 
We will focus on the choice $y_{\rm{cut}}\sim 0.1$, since $a)$ this is the value for which perturbation theory should do a good job at describing collider data, and $b)$ this value corresponds to jets that are somewhat
similar to LHC anti-$k_T$ jets (assuming transverse momentum scaling of the form $p_T \sim \sqrt{y_{\rm{cut}} m_H^2}$). Before proceeding further we first quantify the residual dependence of our \nnnlo predictions 
on the 3-(sub)jettiness slicing parameter $\tau_3^{\rm{cut}}$. We present the $\tau_3^{\rm{cut}}$-dependence of the \nnnlo coefficient for $y_{\rm{cut}}=0.1$ in Fig.~\ref{fig:taudep_n3l}. We have normalized the coefficient to the 
total inclusive correction $\Delta \Gamma^{\rm{N3LO}}_{H\rightarrow b\overline{b}}$ at this order.  To illustrate the size of the power corrections we additionally show 
the function $-2.35-0.00289 \; \tau_3^{\rm{cut}} \ln^3{(\tau_3^{\rm{cut}}/m_H)}$ in the plot. 
 We observe that the $\tau_3^{\rm{cut}}$-dependence for this jet clustering is not dramatic, only changing ~10\% over the range $[0.02-0.3]$ GeV. The dependence between $\tau_3^{\rm{cut}} \sim 0.02-0.05$ GeV is around one percent. Our differential predictions obtained at this order have MC uncertainties around a few percent (on the \nnnlo coefficient) and therefore our results are insensitive to $\tau_3^{\rm{cut}}$ when $\tau_3^{\rm{cut}} \le 0.03$ GeV. We predominately use $\tau_3^{\rm{cut}} =0.02$ GeV for the subsequent differential predictions in this section (supplemented by additional runs with $\tau_3^{\rm{cut}} =0.03$ GeV to improve MC uncertainties in some distributions) . The two-jet rate is around a factor of $-2$ times the inclusive correction at this order, illustrating that there is a large cancellation at this order across jet bins and reminding us that, when exclusive jet quantities are considered, the smallness of an inclusive correction does not necessarily transfer to all distributions and all regions of phase space.

\begin{figure}
\begin{center}
\includegraphics[width=7.5cm]{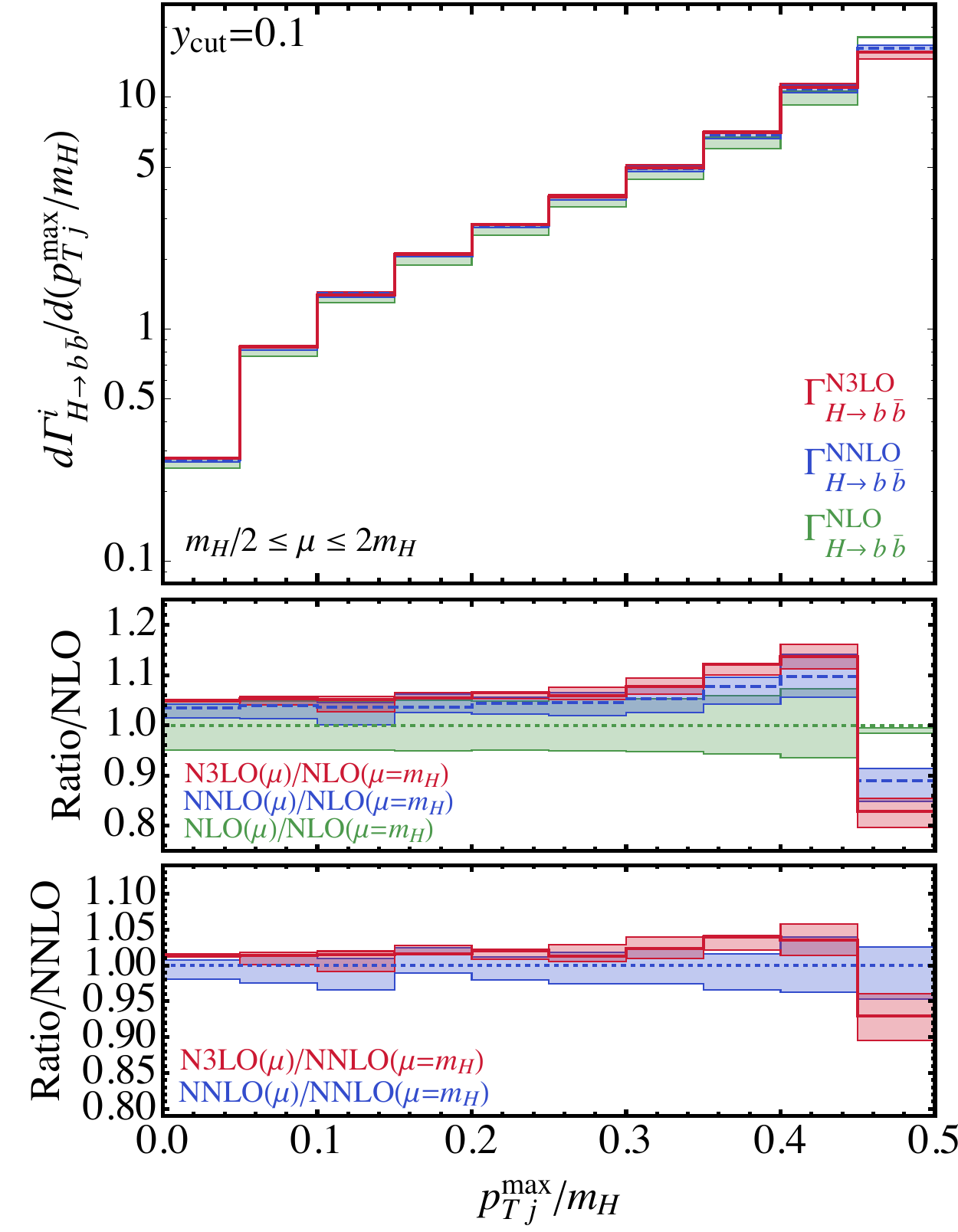}
\includegraphics[width=7.5cm]{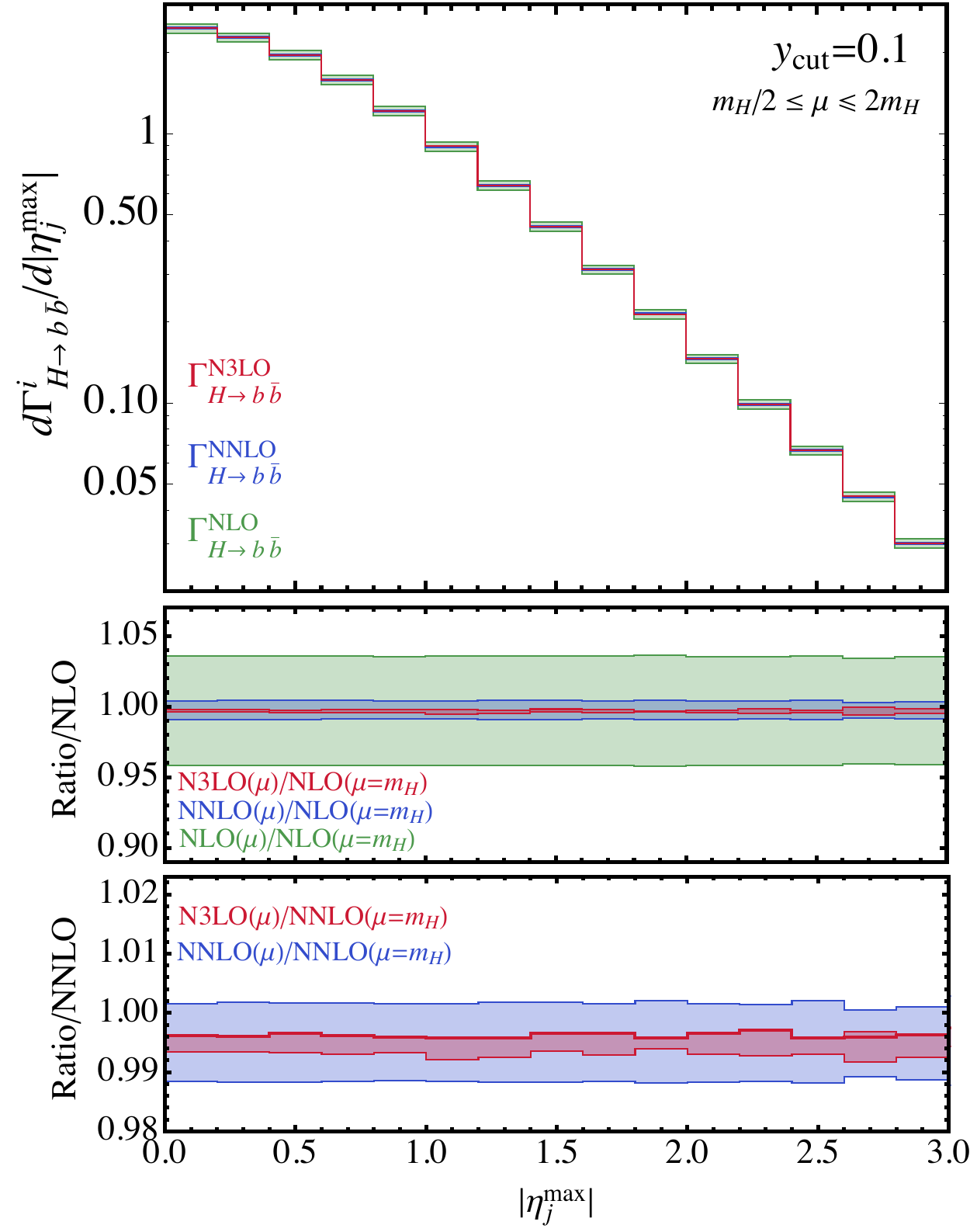}
\caption{The ($m_H$-scaled) transverse momentum and pseudo-rapidity of the maximum-energy jet in the Higgs rest frame at NLO, NNLO, and N$^3$LO.} 
\label{fig:ptrap_n3lo}
\end{center}
\end{figure}

Our final state consists of two jets clustered with the Durham jet algorithm. We distinguish the two jets based upon which has the largest energy component (and refer to them as the max and min jets hereafter). 
As discussed previously, the dynamics of the rest-frame observables is somewhat limited, since physically-relevant distributions such as the energy of the jet and the mass of the jet are delta functions at LO. Therefore, higher-order corrections factorize onto corrections to LO observables $\mathcal{O}_{\rm{LO}}$ which contain contributions from every phase-space region and to observables $\mathcal{O} \ne \mathcal{O}_{\rm{LO}}$ which contain (at most) corrections from one order lower and lack of the two-body phase space. This restricts the ability to study the delicate cancellations that must occur at N$^3$LO. To overcome this, we reintroduce the fictitious collision axis of Section~\ref{sec:IR}, and assume that the $z$-direction is special and corresponds to a beam axis. We then measure the transverse momentum $p_T$ and pseudo-rapidity $\eta$ with respect to this axis. 
This defines non-trivial observables at LO, allowing us to test our predictions more stringently. These predictions also confirm that we can compute jet observables relevant for LHC physics (i.e.~if desired we could impose phase-space cuts on these observables). 

Our results for $|\eta_j^{\rm{max}}|$ and $p_{T,j}^{\rm{max}}/m_H$ are shown in Fig.~\ref{fig:ptrap_n3lo}. We present the NLO, NNLO, and \nnnlo predictions (suppressing LO for clarity).  In each case the upper panel 
presents the differential distribution, while the middle panel illustrates the ratio to the NLO prediction and the lower panel the ratio to the NNLO prediction. Since a scalar particle at rest decays isotropically, the rapidity distribution is sculpted only by the phase-space integration of the final-state jets. For this reason the higher-order corrections are flat and do not noticeably alter the shape of the distribution. As the order
in perturbation theory increases, the scale variation drops considerably (we vary the scale between $m_H/2 \le \mu \le  2 m_H$). This observable inherits the scale variation from the total jet rate and is similar to the 
scale variation presented in Appendix~\ref{sec:Incw} for the total width. At NLO the scale variation is around $\{+3.5, -5\}$\% across the entire distribution. For NNLO and \nnnlo the rate obtained with the scale choice $\mu = m_H$ is close to the maximum rate (again as in the inclusive rate in Appendix~\ref{sec:Incw}), and as such the scale variation band is set by $\mu=m_H$ and $\mu = m_H/2$. At NNLO the variation is around $-1.2\%$ and at \nnnlo it drops by a factor of two to around $-0.7\%$. The $p_T$ distribution is more dynamic, especially in the region $p_T \sim m_H/2$. Here the kinematics of the region is sensitive to the emission of additional soft radiation and thus experiences sizable corrections in the perturbative expansion. At NLO for $p_T \sim m_H/2$ an artificial cancellation of the scale dependence occurs, resulting in essentially no scale dependence in this bin at this order. As the order increases to NNLO and \nnnlo the corrections are around $-10\%$ and $-15\%$ compared to NLO. Across the remaining phase space the corrections are positive and between 5\% in the softest bin 
increasing to around 15\% in the penultimate bin. Comparing \nnnlo to NNLO in the lower panel we see that the \nnnlo corrections reside at the very edge of the scale variation band at NNLO, which corresponds to 
around a 2\% to 5\% correction to the NNLO rate in the bulk region and $-8\%$ correction in the $p_T \sim m_H/2$ bin. This bin has the largest scale variation at \nnnlo corresponding to around $\pm 4\%$. Away from this bin the scale variation at \nnnlo is much smaller, around 1\%. 
 
\begin{figure}
\begin{center}
\includegraphics[width=8cm]{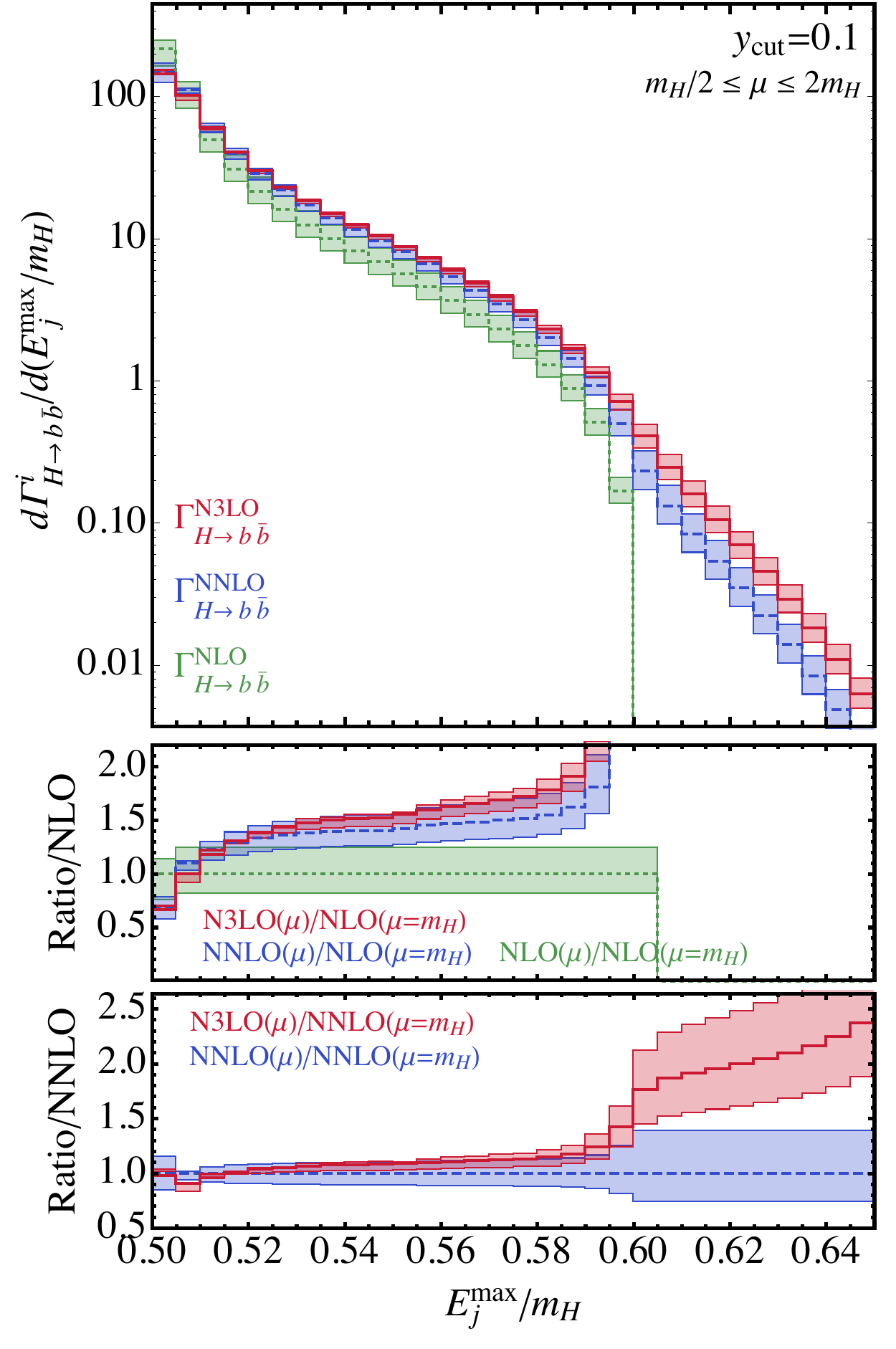}
\caption{The energy component of the four-vector for the jet with maximum energy rescaled by the Higgs mass in the Higgs rest frame at NLO, NNLO, and N$^3$LO.} 
\label{fig:EmaxN3}
\end{center}
\end{figure}

We now turn our attention to the more physically-relevant observables that do not require the introduction of an arbitrary reference direction, namely the energy and invariant mass of the maximum-energy jet. Our results for the ($m_H$-rescaled) energy distribution are presented in Fig.~\ref{fig:EmaxN3}. This observable can broadly be classified into three regions: the $\delta$-component defined by the LO phase space at $E^{\rm{max}}_j = m_H/2$, the ``bulk'' region defined by $ 0.5 <E^{\rm{max}}_j/m_H < 0.6$, and the ``tail'' defined by $E^{\rm{max}}_j > 0.6\,m_H$.  We discuss the  $\delta$-component first, which corresponds to the first bin of our histogram. As can be seen from the middle and lower panels, there is a large (negative) correction in going from NLO to NNLO $(\sim -30\%)$, while the correction in going from NNLO to \nnnlo is much smaller (around $-2\%$), indicating a good convergence of the perturbation series here. The major change in this region at \nnnlo is the dramatic reduction in scale variation compared to NNLO, which has gone from $\pm 15$\% to $+3\%$.  In the bulk region the observable is one order lower in the perturbation theory, i.e.~NLO behaves like LO etc. In our case the \nnnlo correction acts like a NNLO calculation, with the scale variation growing as a function of $E^{\rm{max}}_j$ from a few percent at the softer end to around $10-15\%$ at the more energetic range of the region. The tail region corresponds to a region of phase space which is inaccessible to two- and three-parton phase-space configurations. Therefore in this region the observable behaves like a calculation two orders lower in perturbation theory. As such, the NNLO calculation becomes LO-like (the scale variation in the tail at NNLO is flat since we are merely comparing  the overall factor $m_b^2(\mu_i)\, \alpha^2_s(\mu_i)$ with $\mu_i = \{1/2, 1, 2\}\,m_H$). Since the observable is ``LO'',  we see large corrections $> 2$ and large scale dependence in going from NNLO to N$^3$LO. We note that there exists a ``super-tail'' region not shown in the figure in which $E^{\rm{max}}_j >  0. 65\,m_H$. In this region only the five-parton phase space contributes and therefore the \nnnlo prediction behaves like a LO prediction.

\begin{figure}
\begin{center}
\includegraphics[width=8cm]{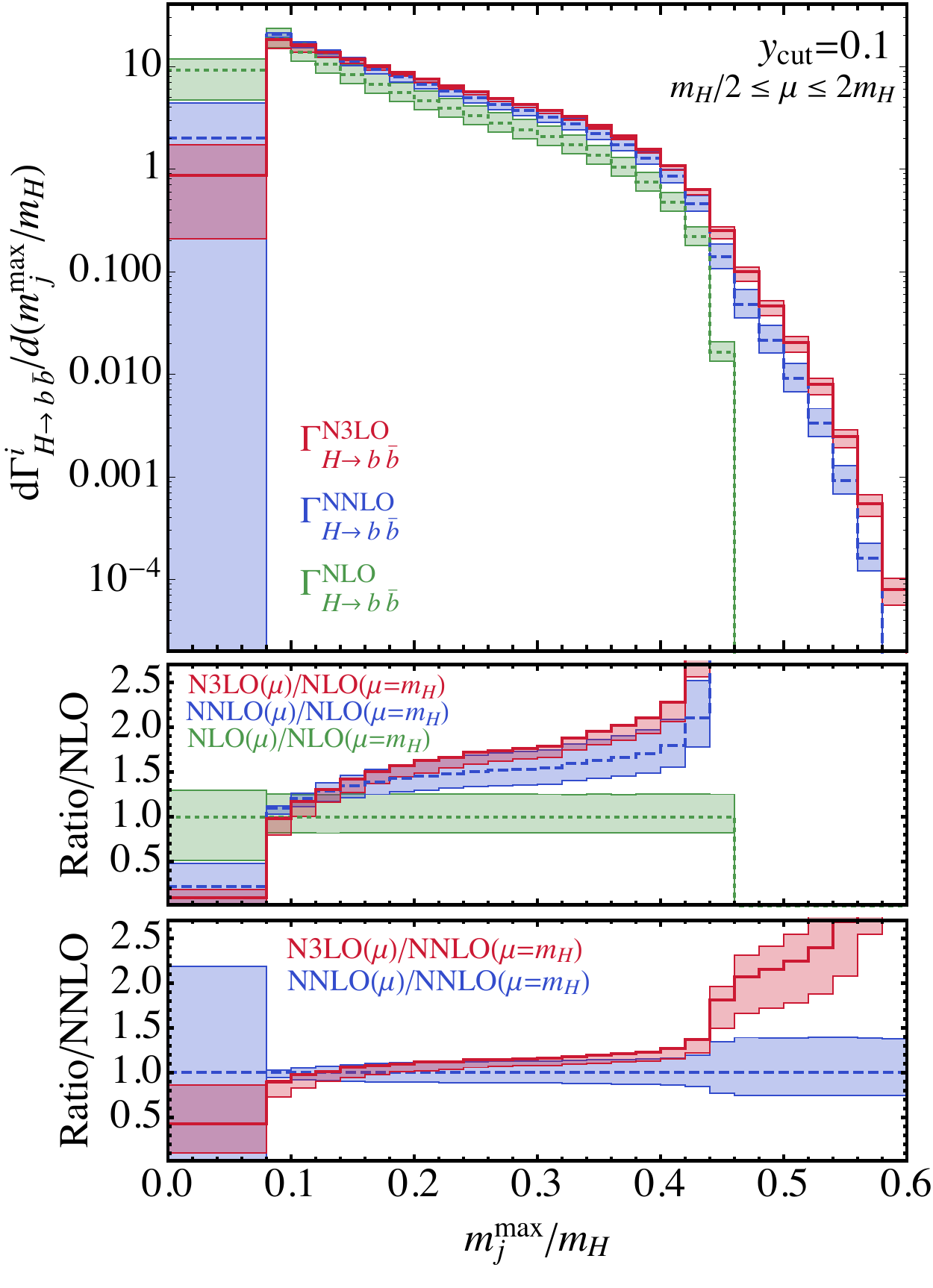}
\caption{The mass of the jet (divided by $m_H$) for the jet with maximum energy in the Higgs rest frame at NLO, NNLO, and N$^3$LO.} 
\label{fig:mj_n3}
\end{center}
\end{figure}

We present the invariant mass of the jet (with the largest energy) $m^j_{\rm{max}}$, divided by the Higgs mass, in Fig.~\ref{fig:mj_n3}. 
At LO all jets are made of single partons and therefore have zero mass\footnote{In the massless approximation. They would have $m_j/m_H \sim 0.02$ had we retained the $b$-quark mass kinematically.}. 
The region near the LO boundary is highly sensitive to soft and collinear radiation, and this observable should be resummed (for instance in a parton-shower prescription) to fully capture the physics. 
In this region of phase space one demands that the most energetic jet be almost massless, which pushes the calculation into the region of phase space in which the two jets are almost-massless partons scattering back to back. In order to obtain a physically-sensible prediction at fixed order one must ensure that the bin near $m_j =0$ is inclusive enough to carry out an adequate cancellation of IR singularities into an IR-safe observable. In other words, if the prediction is binned too finely, the perturbation theory breaks down and undesirable effects (such as a negative differential cross section) can occur. We therefore combine the first four bins into one larger bin in our differential prediction shown in Fig.~\ref{fig:mj_n3}. This is actually insufficient to ensure a physically-reliable prediction for all scale choices at NNLO, but is sufficient at \nnnlo (in which we are primarily interested here). To ensure a positive-definite prediction at NNLO the first five bins need to be combined. We note in passing that at NLO no combination is necessary since the prediction consists only of a three-body phase space (which diverges to $+\infty$ at $\delta(m_j^{\rm{max}}/m_H)$) and of the two-body phase space (which diverges to $-\infty$ at $\delta(m_j^{\rm{max}}/m_H))$). Given the poor convergence of the perturbation series in this region, both higher-order corrections, and the subsequent scale variations, are large. Away from the troublesome $\delta$-region the observable behaves much in the same fashion as the $E_j^{\rm{max}}$ observable discussed previously. Specifically, we observe a bulk region in which the observable is NNLO and the corrections are (reasonably) small and a tail region in which the three-body phase space is not present and the observable becomes NLO, resulting in large corrections at N$^3$LO.

\section{Conclusions}
\label{sec:conc}

In this paper we have presented \nnnlo predictions for the \hbb decay process. We focused on the piece with the most intricate infrared structure, corresponding to diagrams in which the Higgs boson couples directly to the final-state $b\overline{b}$ pair. In order to regulate the IR divergences present at this order we used the Projection-to-Born (P2B) method, employed for the first time with $N$-jettiness slicing as the IR regulator for the NNLO+$j$ contribution. We developed a method of dealing with the requirement of observing a jet direction in the $N$-jettiness slicing approach, namely effectively declustering the last stage of the jet algorithm and using the substructure of the jets to produce three (sub)jet directions. We validated our method at NNLO using three different methods to regulate the IR divergences.  

We used our calculation to present jet rates at $\mathcal{O}(\alpha_s^3)$ and differential distributions for several physical observables using the Durham jet algorithm with $y_{\rm{cut}} = 0.1$. The method discussed in this paper is readily applicable to more complicated Higgs processes, such as associated production of a Higgs boson with a vector boson at the LHC or future collider. We demonstrated this by computing jet observables with respect to an artificial collision axis. Our calculation can also be used outside of the Higgs rest frame. Indeed, since the Higgs is a scalar particle, there is no correlation between decay and production mechanisms. One can therefore always boost any event into the Higgs rest frame, perform the $N$-jettiness regulation (which need not match exactly the requirement of the measurement function, i.e.~one could still employ Durham clustering if desired), then boost back to the laboratory frame and impose additional selection criteria.  We leave this study, together with the inclusion of the remaining top-induced contribution to the \hbb process at $\mathcal{O}(\alpha_s^3)$, to future work.

\acknowledgments

The authors are supported by a National Science Foundation CAREER award number PHY-1652066.
Support provided by the Center for Computational Research at the University at Buffalo.

\appendix

\section{The inclusive \hbb decay width}
\label{sec:Incw}

We present the explicit expressions for the coefficients $s_i$, $\beta_i$ and $\gamma^i_m$ of Eqs.~\eqref{eq:ga1coeff}-\eqref{eq:ga3coeff} following the notation of Ref.~\cite{Chetyrkin:1996sr}. The coefficients $s_i$ read:
\begin{align}
s_{1} &= \frac{17}{4} C_F \\
s_{2} &= \frac{1}{16} \bigg [ C_{F}^{2}\left(\frac{691}{4} - 36\zeta_{2}-36\zeta_{3}\right) + C_{A}C_{F}\left(\frac{893}{4}-22\zeta_{2}-62\zeta_{3}\right) \notag \\ &\quad -C_{F}N_f\left(\frac{65}{2} - 4\zeta_{2}-8\zeta_{3}\right) \bigg ] \\
s_{3} &= \frac{1}{64} \bigg [ C_{F}^{3}\left(\frac{23443}{12} - 648\zeta_{2}-956\zeta_{3}+360\zeta_{5}\right) \notag \\ &\quad + C_{A}C_{F}^{2}\left(\frac{13153}{3}-1532\zeta_{2}-2178\zeta_{3}+580\zeta_{5}\right)  \notag \\ &\quad + C_{A}^{2}C_{F}\left(\frac{3894493}{972}-\frac{6860}{9}\zeta_{2}-\frac{4658}{3}\zeta_{3}+\frac{100}{3}\zeta_{5}\right) \notag \\ &\quad -C_{A}C_{F}N_f\left(\frac{267800}{243}-\frac{2284}{9}\zeta_{2}-\frac{704}{3}\zeta_{3} + \frac{48}{5}\zeta_{2}^{2} - \frac{80}{3}\zeta_{5}\right) \notag \\ &\quad -C_{F}^{2}N_f\left(\frac{2816}{3}-260\zeta_{2}-520\zeta_{3}-\frac{48}{5}\zeta_{2}^{2}+160\zeta_{5}\right)  \notag \\ &\quad +C_{F}N_f^{2}\left(\frac{15511}{243} - \frac{176}{9}\zeta_{2}-16\zeta_{3}\right) \bigg ]
\end{align}
with $C_A=N_c$, $C_F=\frac{N_c^2-1}{2 N_c}$, and $N_f$ the number of quark flavors. The coefficients of the QCD $\beta$ function explicitly read:
\begin{eqnarray}
\beta_{0} &=& \frac{1}{4}\left[\frac{11}{3}C_{A} - \frac{4}{3}T_R N_{f}\right] \\
\beta_{1} &=& \frac{1}{16}\left[\frac{34}{3} C_A^2 - \frac{20}{3} C_A T_R N_f - 4 C_F T_R N_f\right] \,
\end{eqnarray}
with $T_R=\frac{1}{2}$. The coefficients $\gamma^i_m$ are taken from Eq.~(12) of Ref.~\cite{Vermaseren:1997fq} and their expressions are:
\begin{align}
\gamma_{m}^{0} &= \frac{3}{4} C_F \\
\gamma_{m}^{1} &= \frac{1}{16} \left[\frac{3}{2}C_F^2+\frac{97}{6} C_F C_A -\frac{10}{3} C_F T_R N_f \right] \notag \\
\gamma_{m}^{2} &= \frac{1}{64} \bigg [ \frac{129}{2} C_F^3 - \frac{129}{4}C_F^2 C_A + \frac{11413}{108}C_F C_A^2 \notag \\ &\quad +C_F^2 T_R N_f \left(-46+48\zeta_3 \right)
+C_F C_A T_R N_f \left( -\frac{556}{27}-48\zeta_3 \right) \notag \\ &\quad -\frac{140}{27} C_F T_R^2 N_f^2  \bigg ] \, .
\end{align}

\begin{figure}
\begin{center}
\includegraphics[width=9cm]{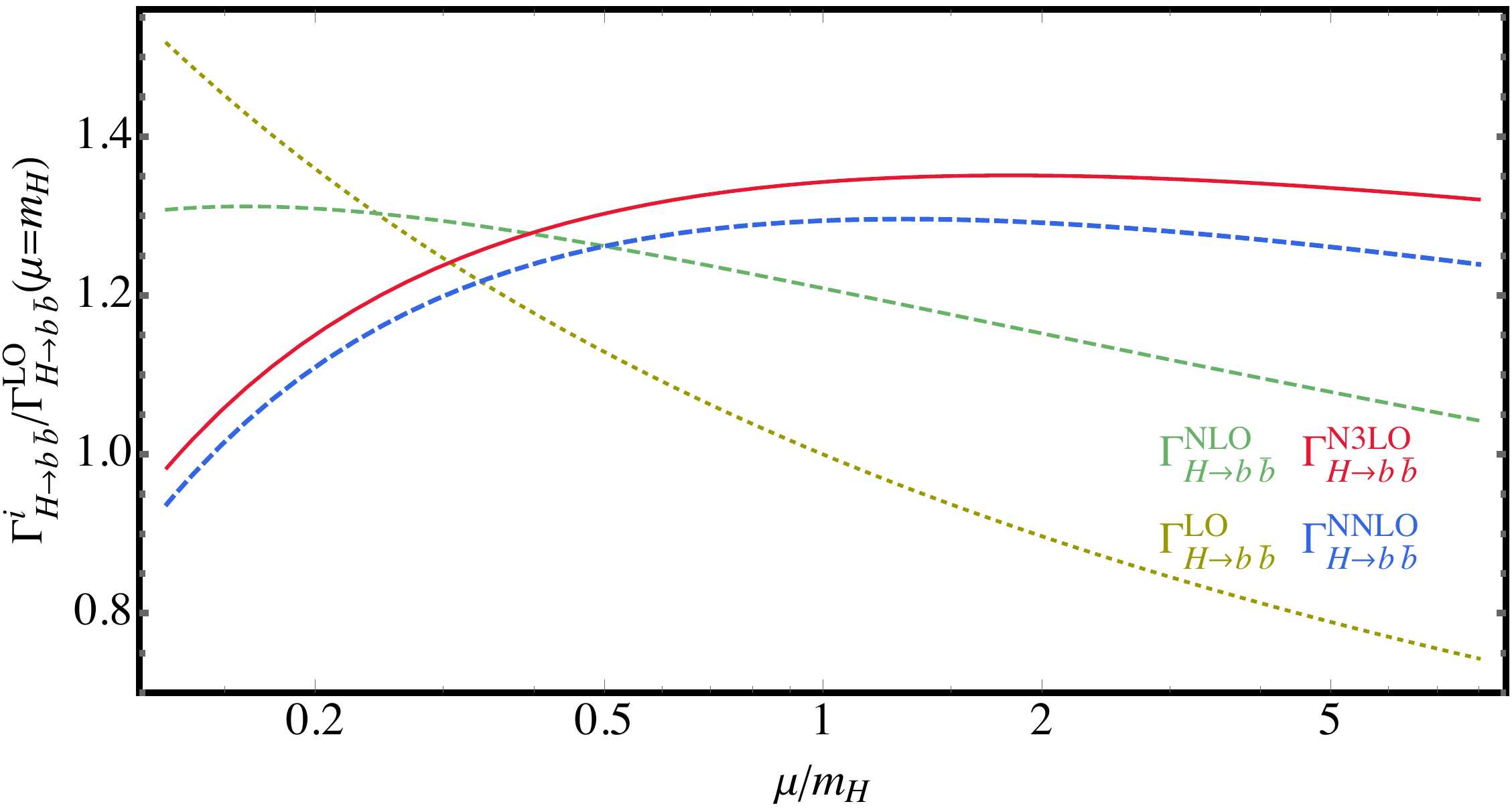}
\caption{Dependence on the renormalization scale $\mu$ of the inclusive \hbb decay width up to N$^3$LO accuracy (rescaled by the LO width at $\mu=m_H$).} 
\label{fig:IncRate}
\end{center}
\end{figure}

Finally, it is instructive to show the renormalization scale variation of the inclusive \hbb  decay width up to $\mathcal{O}(\alpha_s^3)$. The inclusive decay width depends on the renormalization scale $\mu$ through the bottom Yukawa coupling $y_b(\mu)$, the strong coupling constant $\alpha_{s}(\mu)$, and the coefficients $\Gamma^{(n)}_{H\rightarrow b\overline{b}}$ of Eqs.~\eqref{eq:ga1coeff}-\eqref{eq:ga3coeff}. We plot the ratios $\Gamma^{\text{N}^n\text{LO}}_{H\rightarrow b\overline{b}}(\mu)/\Gamma^{\text{LO}}_{H\rightarrow b\overline{b}}(\mu=m_H)$ with $n=0,\dots,3$ as $\mu/m_H$ is varied in the range $\{1/8,8\}$ in Fig.~\ref{fig:IncRate}. The values of $\alpha_s$ and $y_b$ at different scales are obtained using the Mathematica package RunDec \cite{Chetyrkin:2000yt}. As expected, the inclusion of higher-order corrections stabilizes the inclusive decay width, which shows very small scale dependence at N$^3$LO in the primary region of interest $\{1/2,2\}m_H$.

\bibliographystyle{JHEP}

\bibliography{HbbN3LO}

\end{document}